\documentclass[12pt,preprint]{emulateapj}
\usepackage{lscape,url,color}
\usepackage{amsmath}
\usepackage{relsize}

\shorttitle{Luhman16AB: Clouds, Temperature, and Gravity.}
\shortauthors{Faherty et al.}
\begin{document}

\title{Signatures of Cloud, Temperature, and Gravity From Spectra of the Closest Brown Dwarfs\footnote{This paper includes data gathered with the 6.5 meter Magellan Telescopes located at Las Campanas Observatory, Chile.}}

\author{Jacqueline K.\ Faherty\altaffilmark{1,2,3,9}, Yuri Beletsky\altaffilmark{4}, Adam J.\ Burgasser\altaffilmark{5,8}, Chris Tinney\altaffilmark{6}, David J.\ Osip\altaffilmark{4},  Joseph C. Filippazzo\altaffilmark{2,10}, Robert A.\ Simcoe\altaffilmark{7}}

\altaffiltext{1}{Department of Terrestrial Magnetism, Carnegie Institution of Washington, Washington, DC 20015, USA; jfaherty@ciw.edu }
\altaffiltext{2}{Department of Astrophysics, 
American Museum of Natural History, Central Park West at 79th Street, New York, NY 10034 }
\altaffiltext{3}{Department of Astronomy, Universidad de Chile Cerro Calan, Las Condes, Chile}
\altaffiltext{4}{Las Campanas Observatory, Carnegie Institution of Washington, Colina el Pino, Casilla 601 La Serena, Chile}
\altaffiltext{5}{Center of Astrophysics and Space Sciences, Department of Physics, University of California, San Diego, CA 92093, USA}
\altaffiltext{6}{University of New South Wales Sydney, Australia}
\altaffiltext{7}{MIT-Kavli Institute for Astrophysics and Space Research, 70 Vassar Street, Cambridge, MA 02139}
\altaffiltext{8}{Hellman Fellow}
\altaffiltext{9}{Hubble Fellow}
\altaffiltext{10}{Department of Engineering Science and Physics, College of Staten Island, 2800 Victory Boulevard, Staten Island, NY 10301 }

\begin{abstract}
We present medium resolution optical ($\lambda$/$\Delta\lambda$$\sim$4000) and near-infrared ($\lambda$/$\Delta\lambda$$\sim$8000) spectral data for components of the newly discovered  WISE J104915.57-531906.1AB  (Luhman 16AB) brown dwarf binary.    The optical spectra reveal strong 6708\,\AA\  Li I absorption in both Luhman 16A  (8.0$\pm$0.4 \AA) and Luhman 16B (3.8$\pm$0.4 \AA) confirming their substellar mass.  Interestingly, this is the first detection of Li I absorption in a T dwarf.  In the near-infrared data, we find strong K~I absorption  at 1.168, 1.177, 1.243, and 1.254 $\mu$m in both components.  Neither the optical nor the near-infrared alkali lines show low-surface gravity signatures. Along with the Li I absorption detection, we can broadly constrain the system age to 0.1-3 Gyr or the mass to 20 - 65 M$_{Jup}$ for each component. Compared to the strength of K~I line absorption in equivalent spectral subtype brown dwarfs, Luhman 16A is weaker while Luhman 16B is stronger.  Analyzing the spectral region around each doublet in distance scaled flux units and comparing the two sources,  we confirm the $J$ band flux reversal and find that Luhman 16B has a brighter continuum in the 1.17 $\mu$m and 1.25 $\mu$m regions than Luhman 16A.  Converting flux units to a brightness temperature we interpret this to mean that the secondary is $\sim$ 50 K warmer than the primary in regions dominated by condensate grain scattering.  One plausible explanation for this difference is that Luhman 16B has thinner clouds or patchy holes in its atmosphere allowing us to see to deeper, hotter regions.  We also detect comparably strong FeH in the 0.9896 $\mu$m Wing-Ford band for both components.  Traditionally, a signpost of changing atmosphere conditions from late-type L to early  T, the persistence and similarity of FeH at 0.9896 $\mu$m in both Luhman 16A and Luhman 16B is an indication of homogenous atmosphere conditions.   We calculate bolometric luminosities from observed data supplemented with best fit models for longer wavelengths and find the components are consistent within 1$\sigma$ with resultant T$_{effs}$ of 1310$\pm$30 K and 1280$\pm$75 K for Luhman 16AB respectively. 
\end{abstract}

\keywords{ binaries: visual -- stars: individual (WISE J104915.57-531906.1) -- stars: low mass, brown dwarfs}
\section{INTRODUCTION}

Not since the characterization of Wolf 359 in 1928, has the list of the five closest stellar systems to the Sun been altered (\citealt{van-Maanen28}).  That changed with the recent discovery  by \citet{Luhman13} of the brown dwarf binary WISE J104915.57-531906.1AB (Luhman 16AB here-after) at a distance of just 2.02$\pm$0.019pc (\citealt{Boffin13}).  Naturally, the Sun's closest neighbors become observational standards. They are inevitably the best studied astronomical targets and provide detailed information which forms the baseline for our understanding of similar objects. The Luhman 16AB system is not only nearby, and a co-evolving binary (L7.5+T0.5 -- \citealt{Burgasser13}), but it also covers a critical temperature range for our understanding of cool atmospheres.

\begin{figure*}[!t]
\begin{center}
\resizebox{1.0\hsize}{!}{\includegraphics[clip=true]{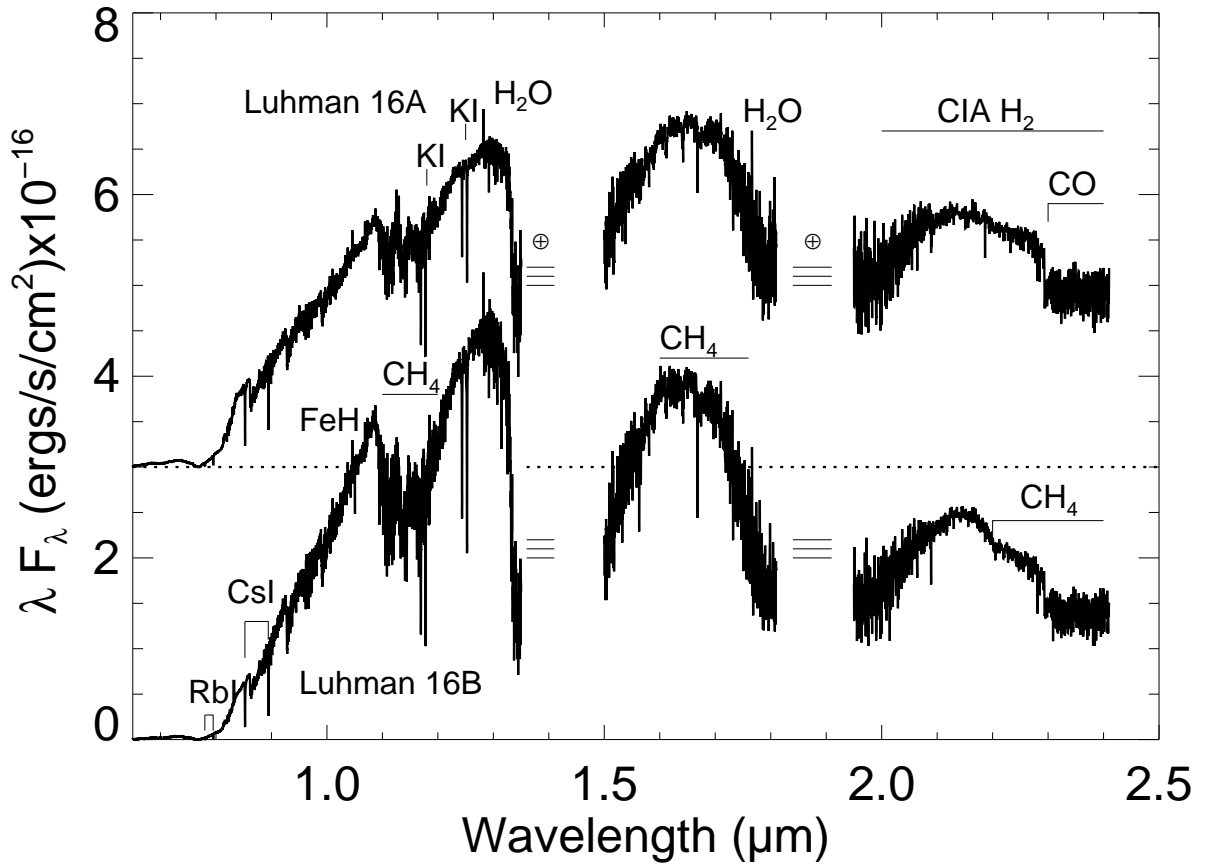}}
\end{center}
\caption{Medium resolution MagE optical ($\lambda$/$\Delta\lambda$$\sim$4000) and FIRE near-infrared ($\lambda$/$\Delta\lambda$$\sim$8000) spectral data for Luhman 16A (L7.5-top) and Luhman 16B (T0-bottom) with prominent features labeled.  Areas of strong telluric absorption at $\sim$1.4 $\mu$m and $\sim$1.9 $\mu$m have been removed but are marked by three horizontal lines.  The two sources are offset from one another by 3.0x10$^{-16}$ units as indicated by the dashed line.   We have used the distance of 2.02$\pm$0.019 pc reported in \citet{Boffin13} and the resolved photometry from \citet{Burgasser13} to scale the data to the inferred absolute flux densitites.  } 
\label{fig:fullnir}
\end{figure*}

\begin{figure*}[!t]
\resizebox{1.0\hsize}{!}{\includegraphics[clip=true]{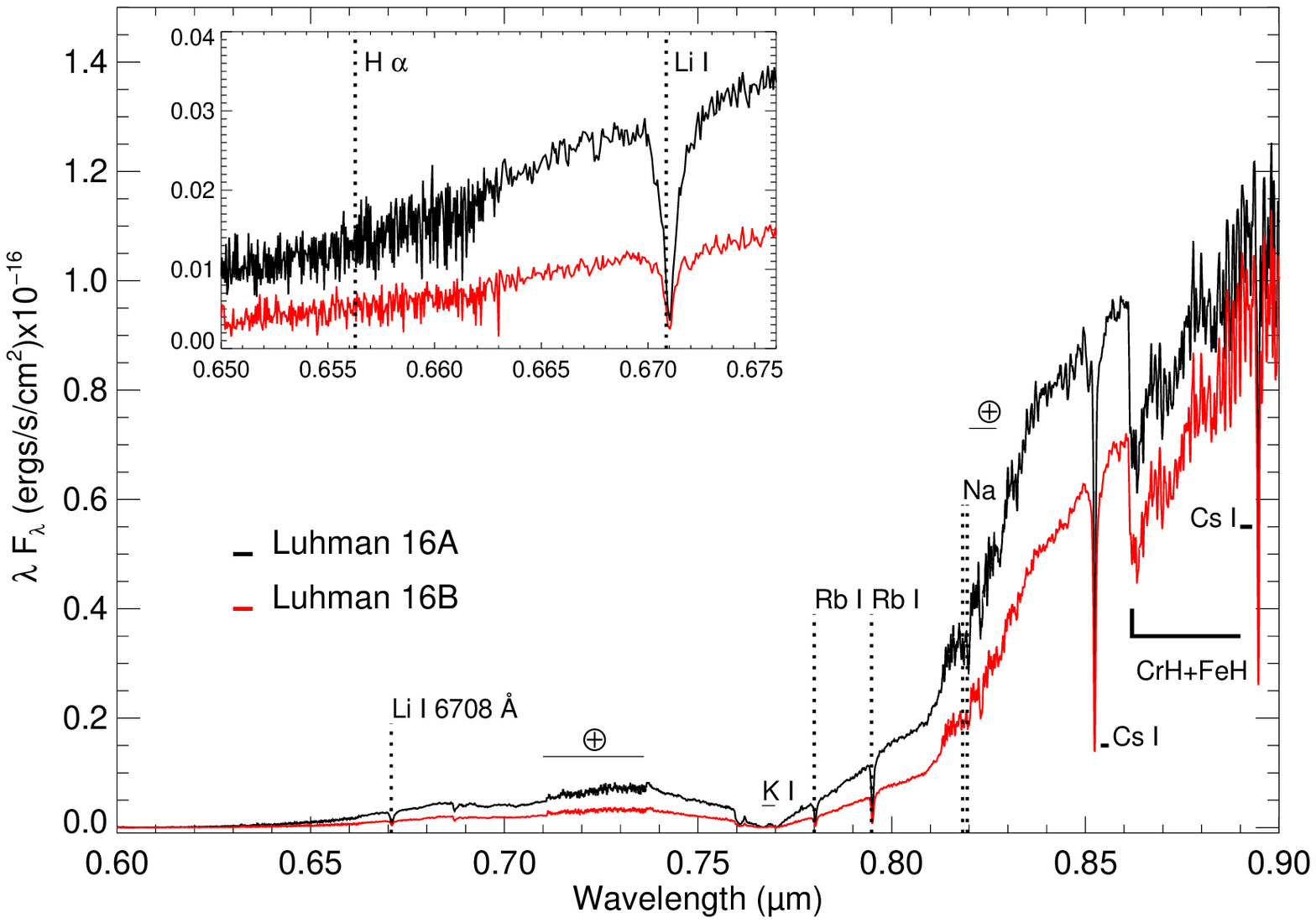}}
\caption{The MagE optical ($\lambda$/$\Delta\lambda$$\sim$ 4000) data for Luhman 16A (black--upper) and Luhman 16B (red--lower) scaled to the distance of 2.02$\pm$0.019 pc reported in \citet{Boffin13}. Prominent features are labeled.  No offset has been applied between components and the spectrum has not been telluric corrected.  Highlighted in the inset box at top left is the region around the 6708\,\AA\  Li I absorption line.  } 
\label{fig:Mage}
\end{figure*}

Brown dwarf observables are shaped by gas and condensation chemistry.  Their low temperatures and high-pressures (1 bar $<$ P $<$ 10 bar) favor the formation of molecules such as CO, CH$_{4}$, N$_{2}$, NH$_{3}$, and H$_{2}$O.    For warmer brown dwarfs (e.g. L dwarfs),  both liquid (e.g. Fe) and solid (e.g. CaTiO3, VO) mineral and metal condensates settle into discrete cloud layers (e.g. \citealt{Lodders02}, \citealt{Visscher10}, \citealt{Ackerman01}, \citealt{Marley02}, \citealt{Tsuji02}, \citealt{Woitke04}). As temperatures cool into the T dwarfs, dust clouds form at such deep levels in the photospheres that they have little or no impact on the emergent spectrum.  The transition between ``cloudy" to ``cloudless" objects occurs rapidly over a narrow temperature range (1200-1400 K or L-type into T-type) and drives extreme photometric, spectroscopic, and luminosity changes (\citealt{Burgasser02a, Burgasser08}, \citealt{Tinney03}, \citealt{Vrba04}, \citealt{Golimowski04}, \citealt{Faherty12}, \citealt{Dupuy12}, \citealt{Radigan12, Radigan14}, \citealt{Artigau09}, \citealt{Wilson14}). The mechanism for this cloud-clearing is still hotly debated and may be due to cloud thinning, rain-out, or some combination of the two (\citealt{Ackerman01}, \citealt{Burgasser02}, \citealt{Knapp04}, \citealt{Saumon08}, \citealt{Apai13}, \citealt{Buenzli12}).

Understanding cloud properties and subsequent weather patterns is important for interpreting the observable properties of not only brown dwarfs but planets as well.  Studies of giant planetary mass companions with effective temperatures squarely in the brown dwarf regime have demonstrated that clouds are a critical parameter in shaping directly imaged data (\citealt{Barman11}, \citealt{Marley12}, \citealt{Madhusudhan11}).  The Luhman 16AB binary, which contains the two brightest examples of the L-T transition in an assumed co-evolving system, is poised to become a benchmark source for low-temperature atmosphere studies.  Indeed, recent work has shown that photometric and spectroscopic variations explained by weather patterns on the primary in this system will greatly inform our knowledge of extrasolar planetary atmospheric physics (\citealt{Biller13}, \citealt{Gillon13}, \citealt{Burgasser13,Burgasser13a,Burgasser14}, \citealt{Crossfield14}).

In this paper we show medium resolution optical and near-infrared spectra of both components of Luhman 16AB.  In Section 2 we discuss the data collected for this work.  In Section 3 we break the spectra into individual bandpasses and discuss temperature and gravity indications.  In Section 4 we discuss cloud features revealed in the data.  In Section 5 we conduct a model comparison to examine the quality of fits and resultant fundamental parameters.   Conclusions are presented in Section 6.  

\begin{figure*}[!ht]
\begin{center}$
\begin{array}{cc}

\includegraphics[width=3.5in]{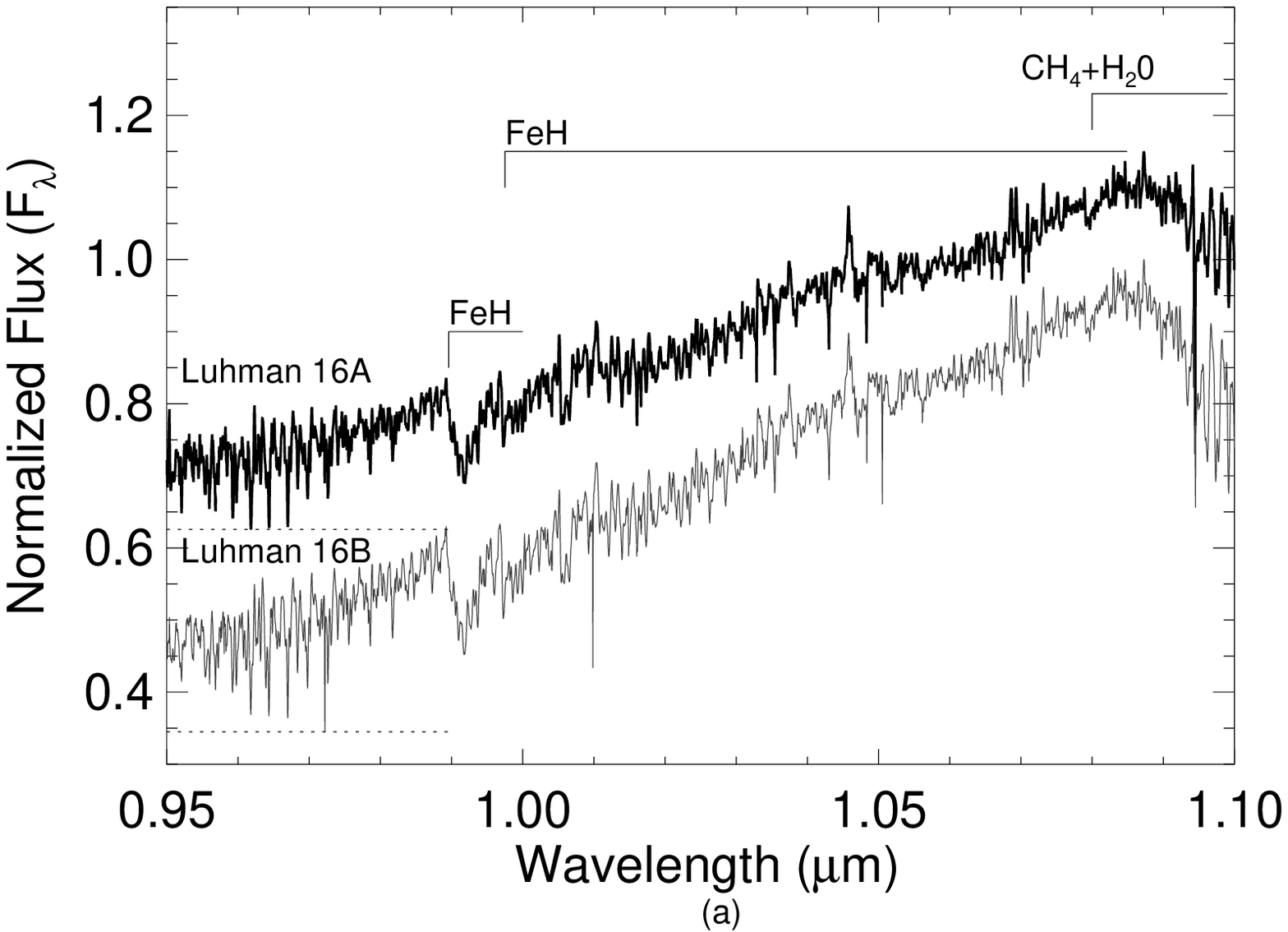}&
\includegraphics[width=3.5in]{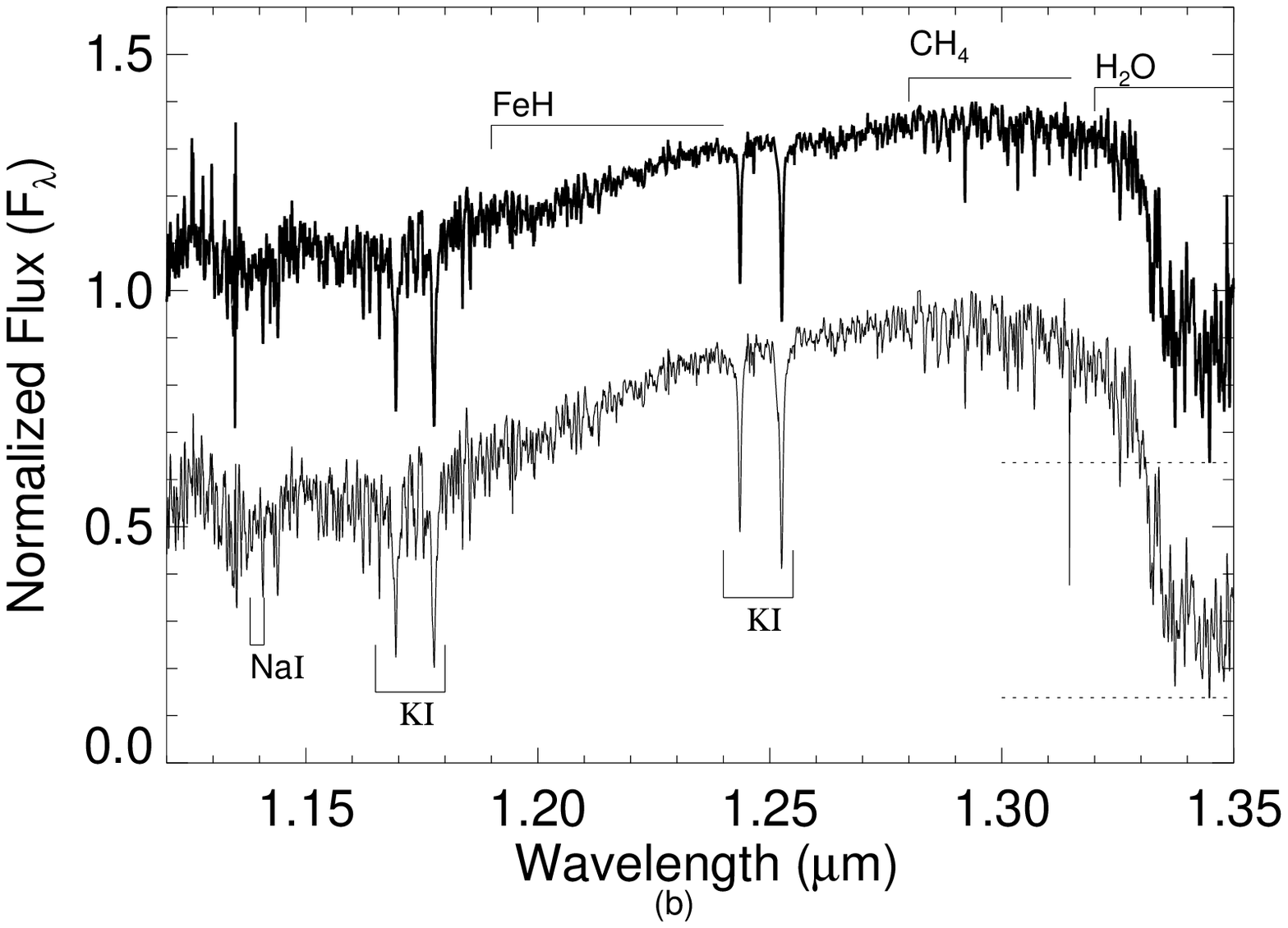} \\
\includegraphics[width=3.5in]{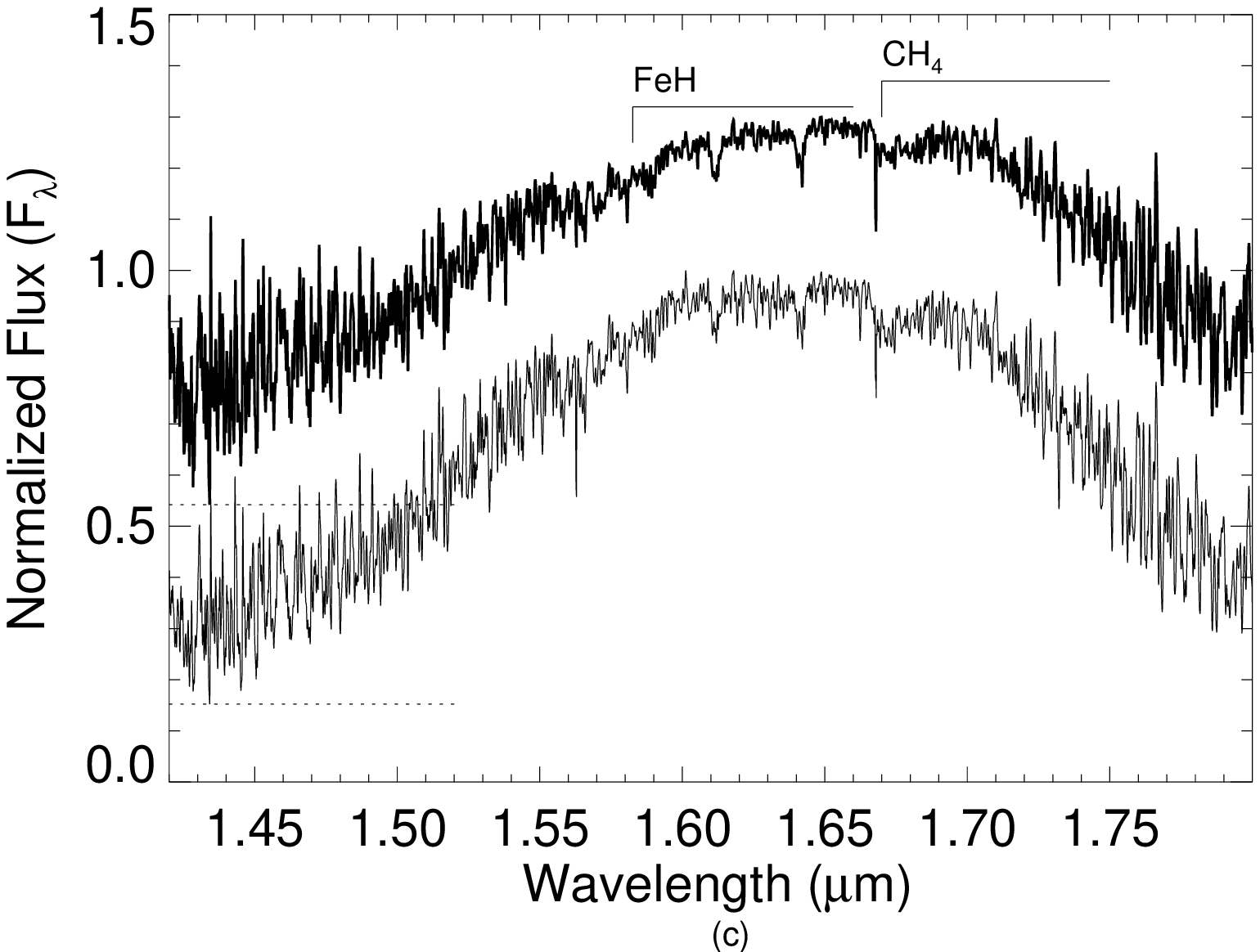} &
\includegraphics[width=3.5in]{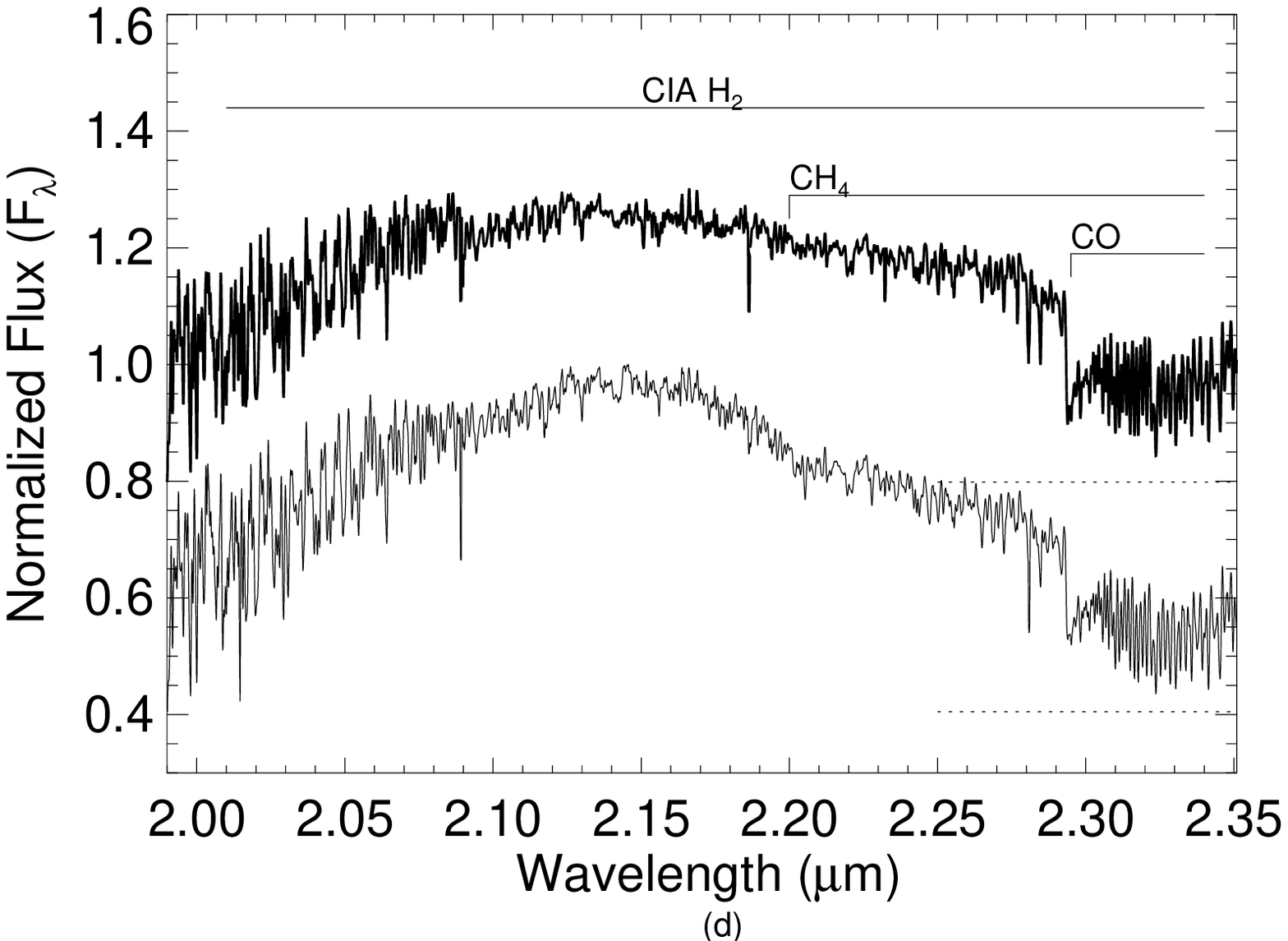} \\

\end{array}$
\end{center}
\caption{The FIRE near-infrared data for Luhman 16A and Luhman 16B normalized over the peak of the region shown of (a) $z$ band, (b) $J$ band, (c) $H$ band, and (d) $K$ band.  In each panel, Luhman 16A is offset from Luhman 16B by a constant (0.15, 0.4, 0.3, and 0.3 for (a), (b), (c), (d) respectively) and shown on top. The minimum flux (in the region shown) for both components is marked by a short dashed line. Prominent features are labeled throughout.  }
\label{fig:spectra-bands}
\end{figure*}

\begin{figure}[ht!]
\begin{center}
\resizebox{1.0\hsize}{!}{\includegraphics[clip=true]{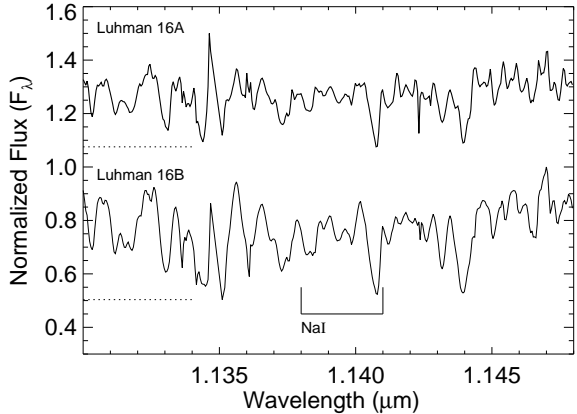}}
\end{center}
\caption{The normalized flux of Luhman 16A (top spectra) and Luhman 16B (bottom spectra) shown around the (1.138, 1.141) $\mu$m Na I  doublet.  Flux is normalized over the peak of the region shown and sources are offset from one another by 0.5. The minimum flux (in the region shown) for both components is marked by a short dashed line.} 
\label{fig:Na I}
\end{figure}

\begin{figure*}[!ht]
\begin{center}$
\begin{array}{cc}
\includegraphics[width=3.5in]{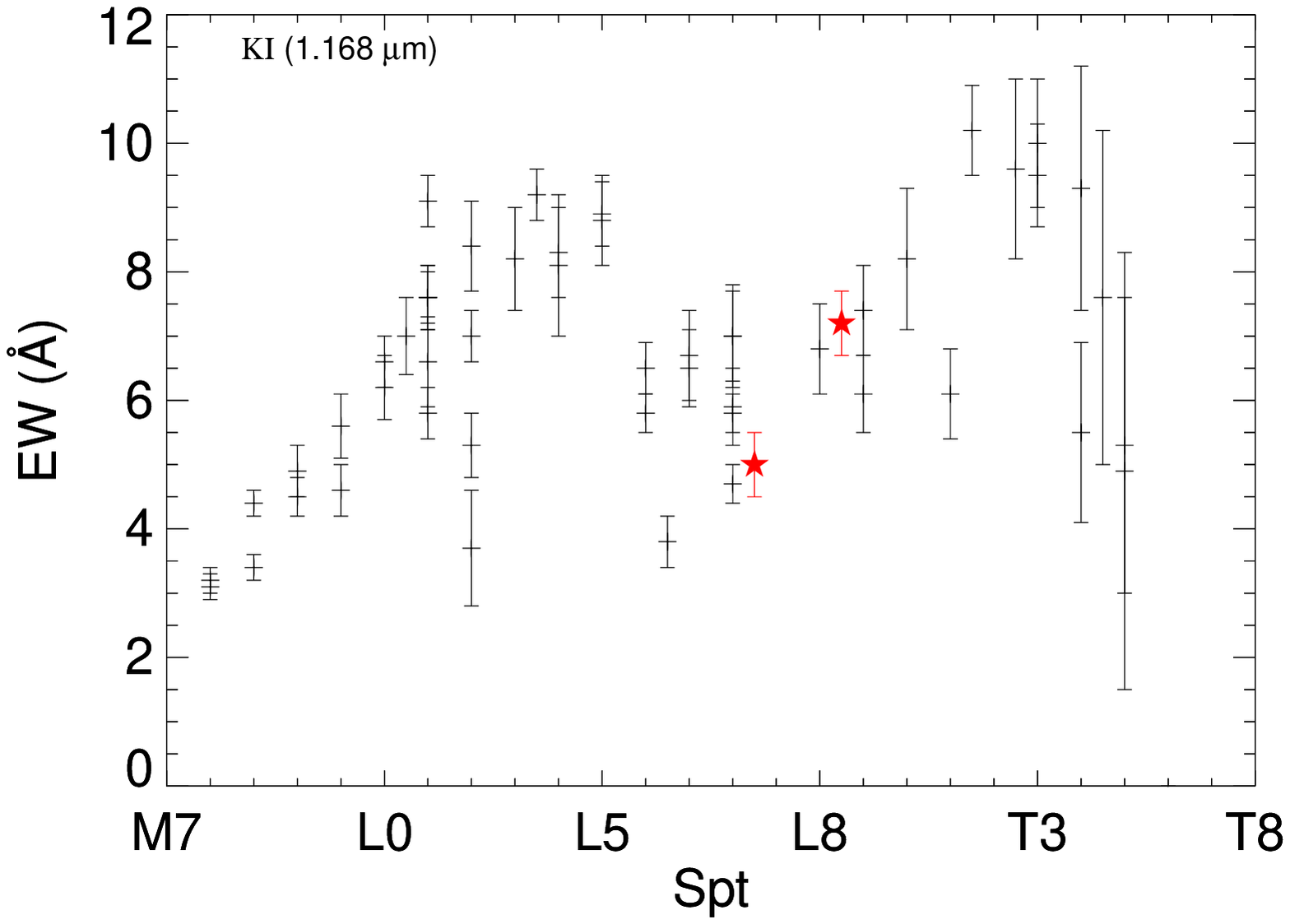}&
\includegraphics[width=3.5in]{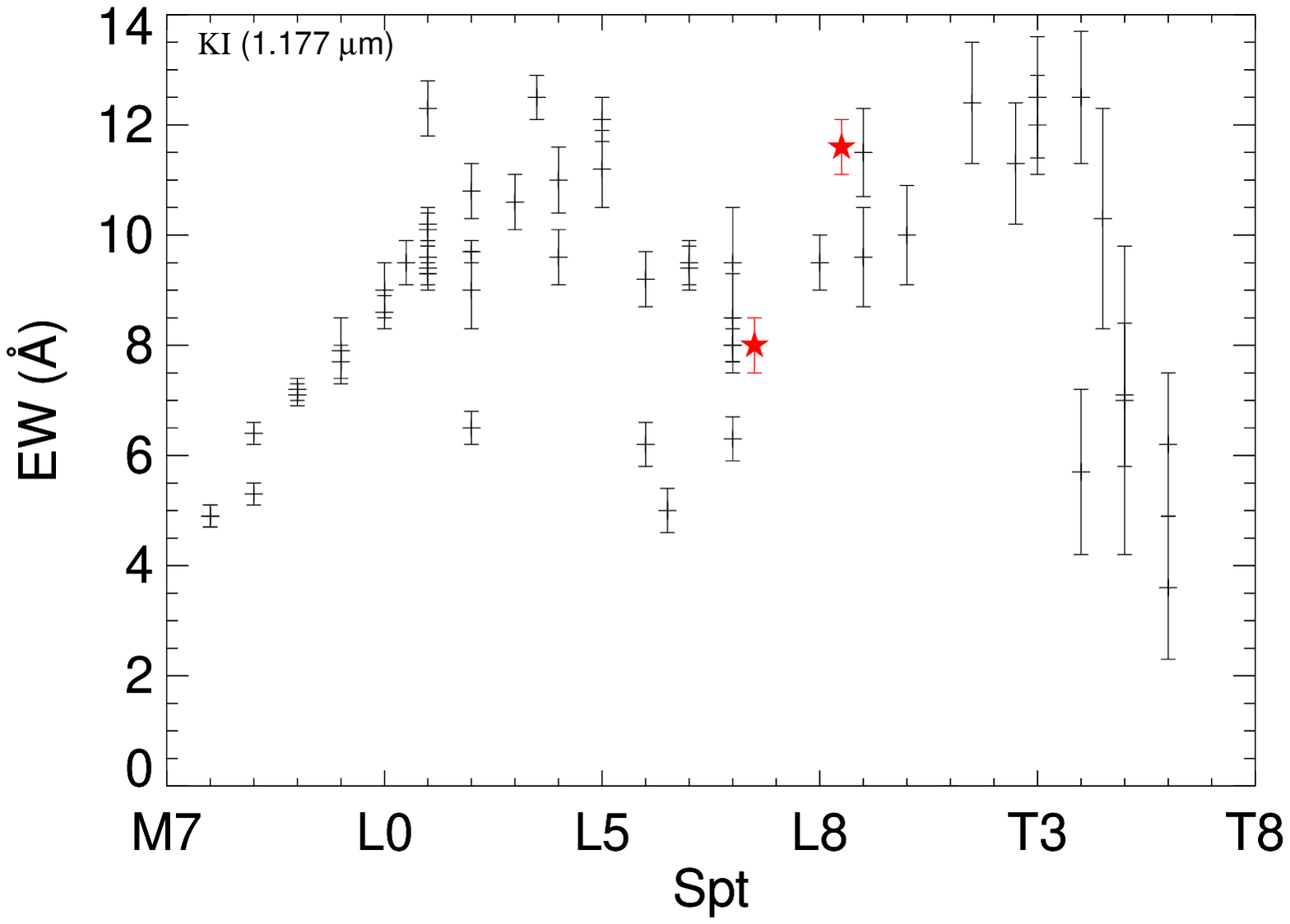}  \\
\includegraphics[width=3.5in]{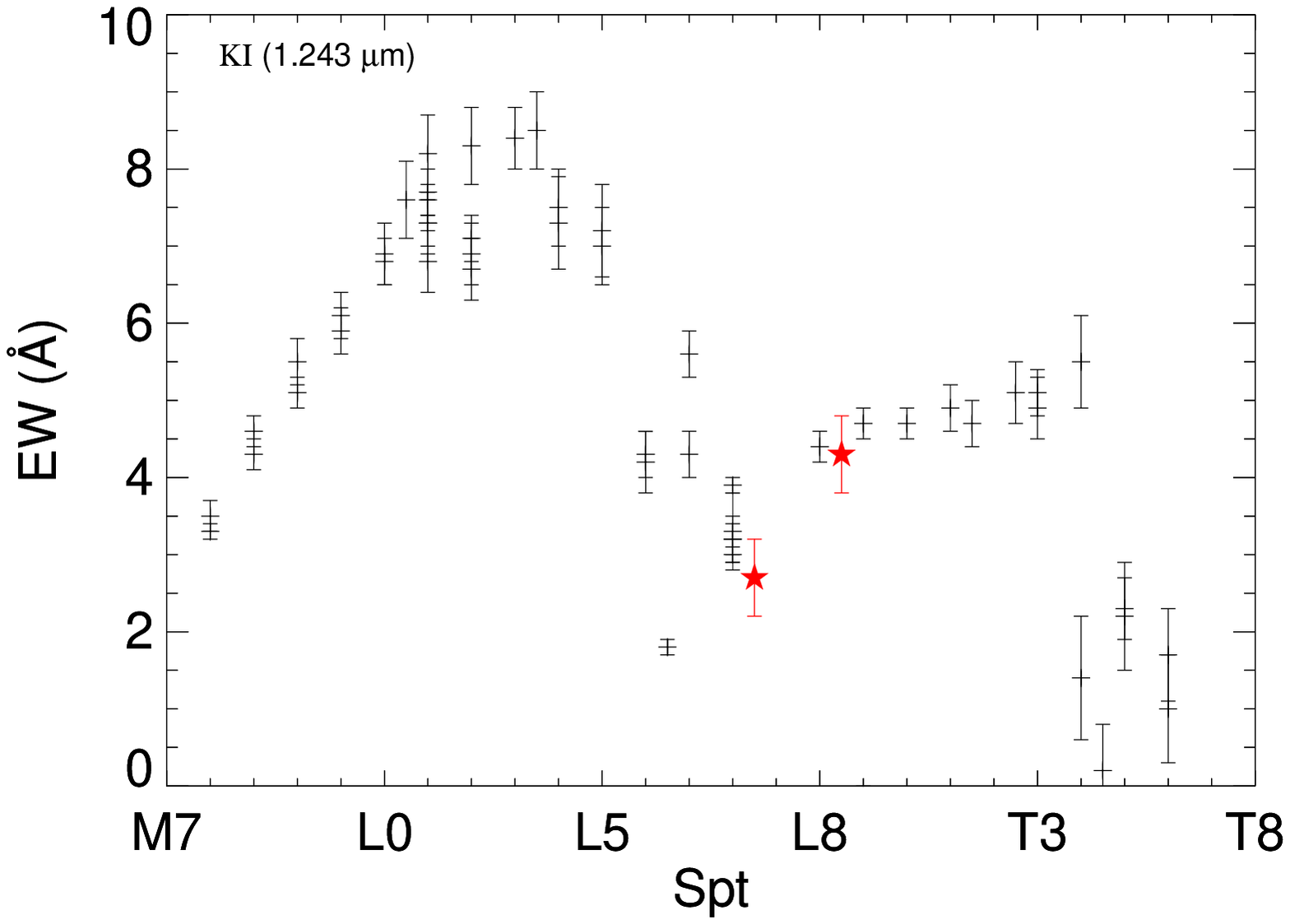}&
\includegraphics[width=3.5in]{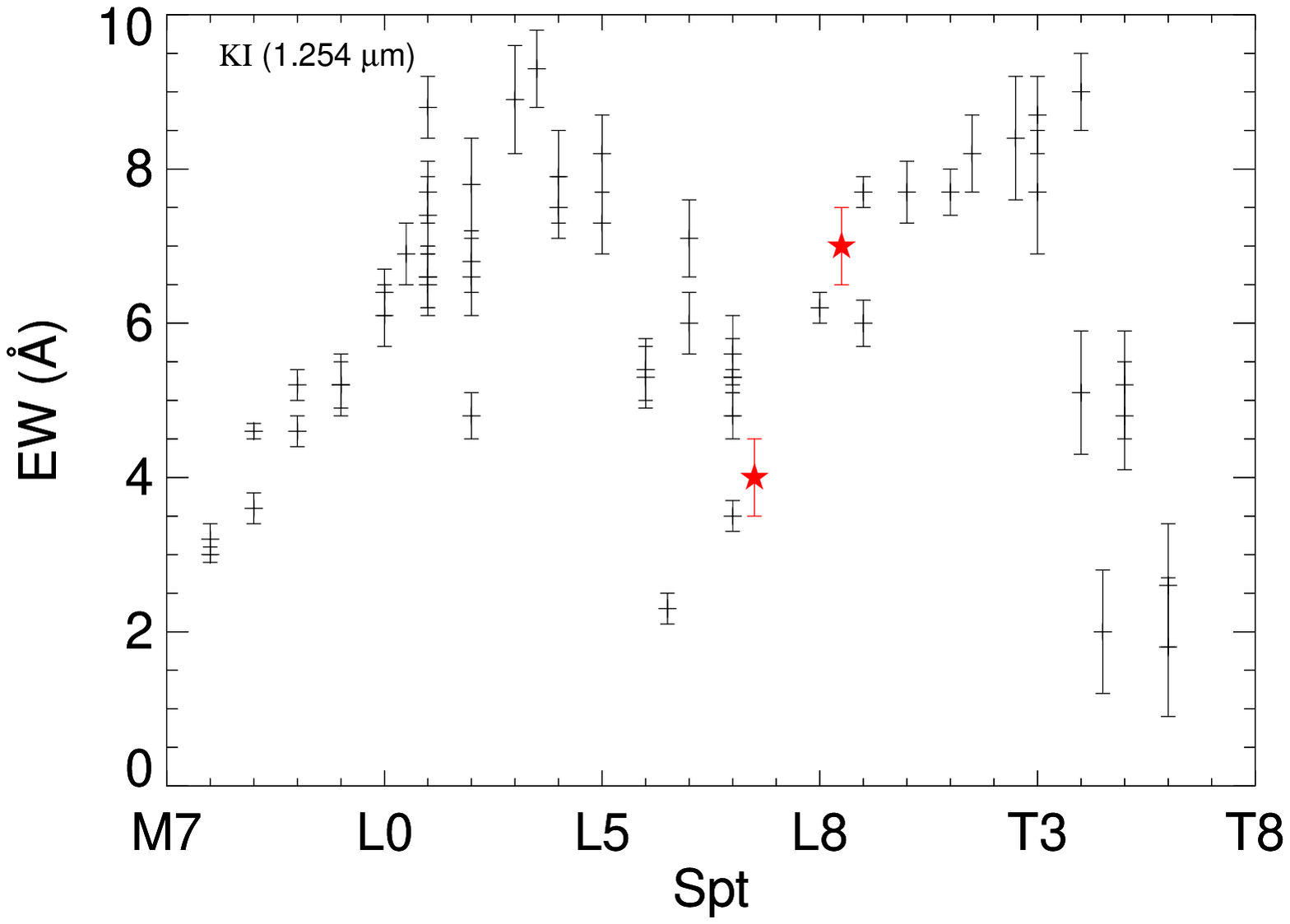}  \\

\end{array}$
\end{center}
\caption{A comparison of the equivalent widths of the K~I 1.168 $\mu$m (top left), 1.177 $\mu$m (top right), 1.243 $\mu$m (bottom left), and 1.254 $\mu$m (bottom right) lines for each component (listed in Table ~\ref{Table:EQ_IR}--marked as red five-pointed star) to the sample of ultra cool dwarfs in \citet{McLean03}. }
\label{fig:mclean}
\end{figure*}

\begin{figure*}[!ht]
\begin{center}$
\begin{array}{cc}
\includegraphics[width=3.5in]{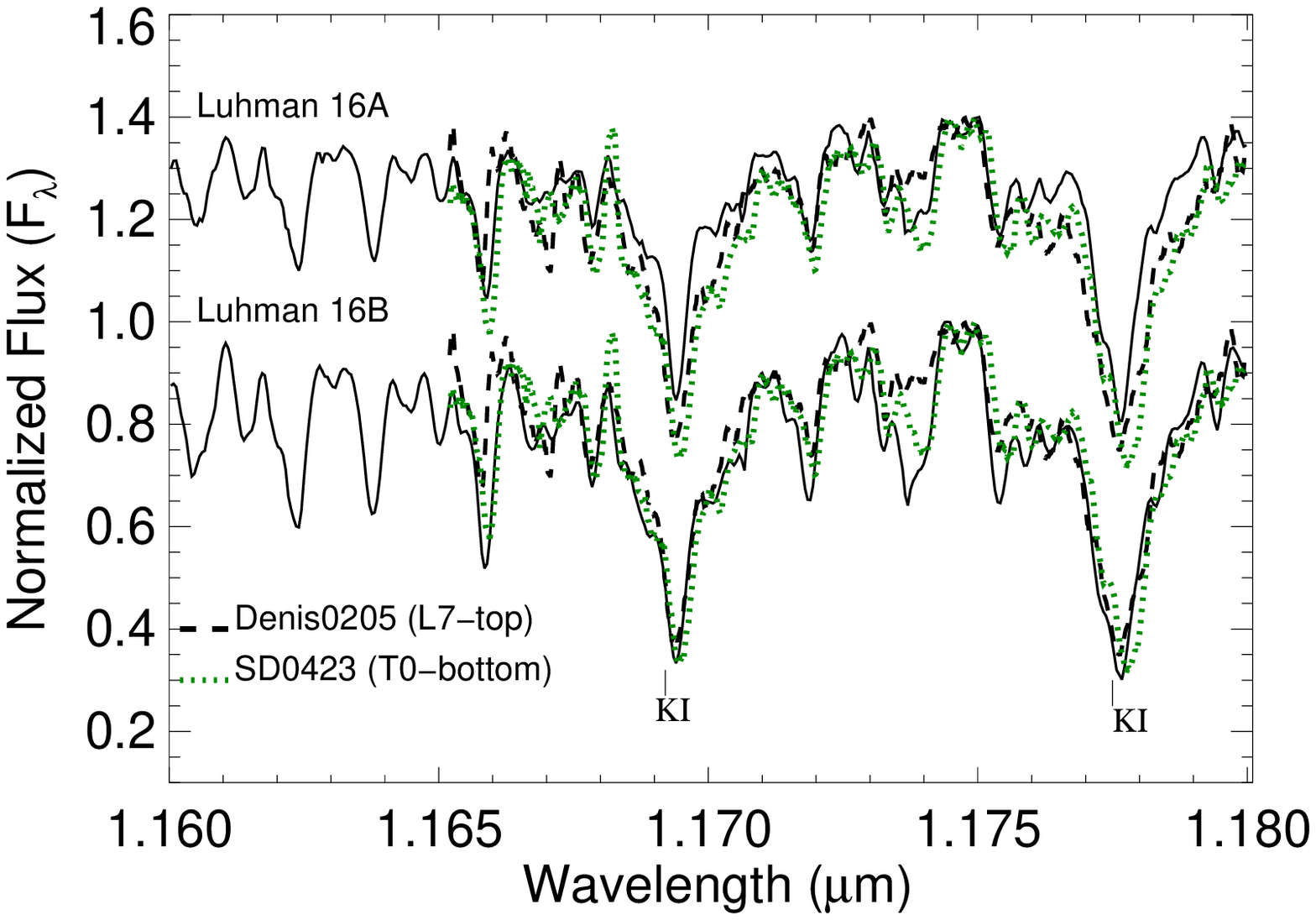}{(a)} &
\includegraphics[width=3.5in]{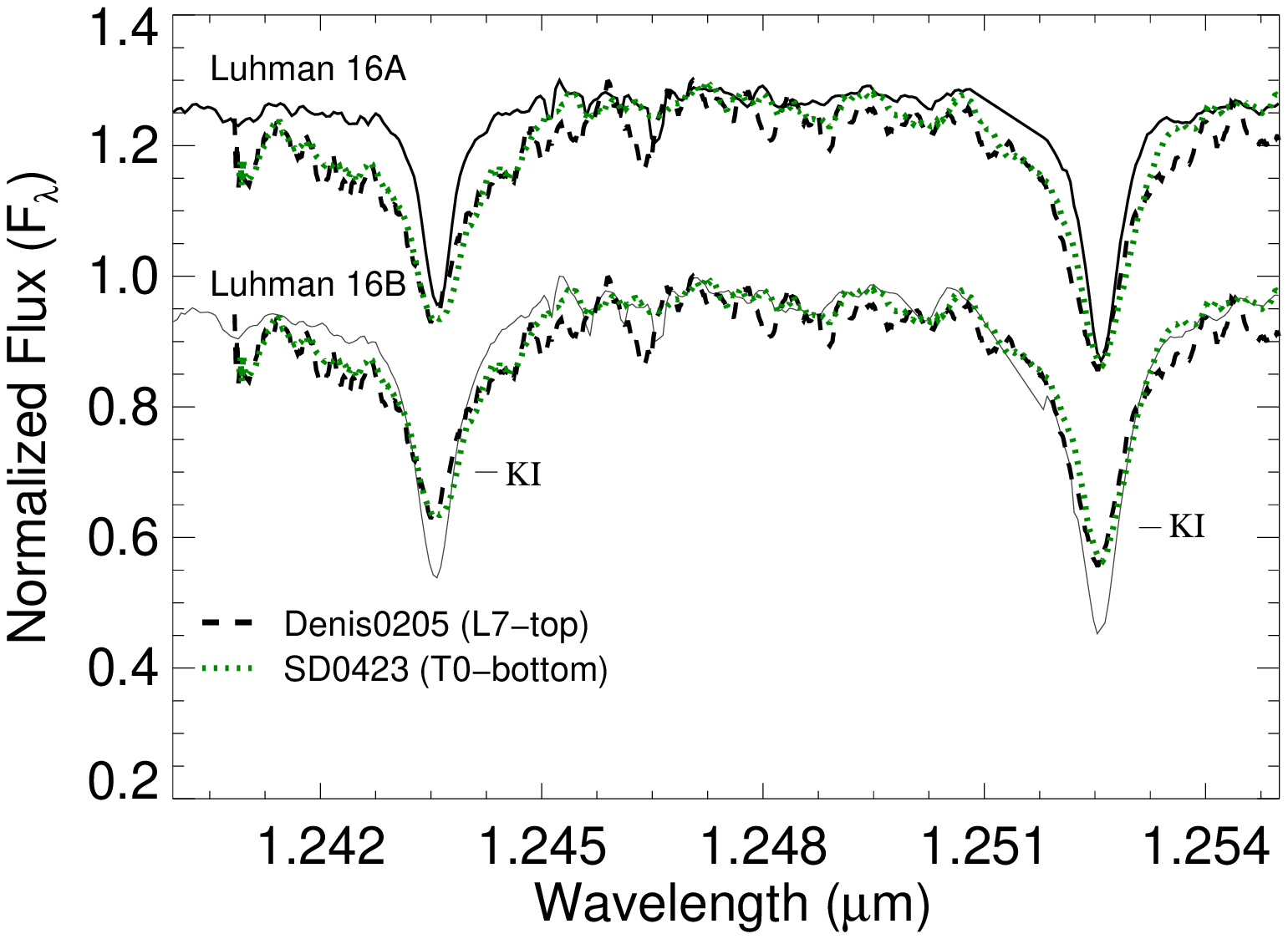} \\
\includegraphics[width=3.5in]{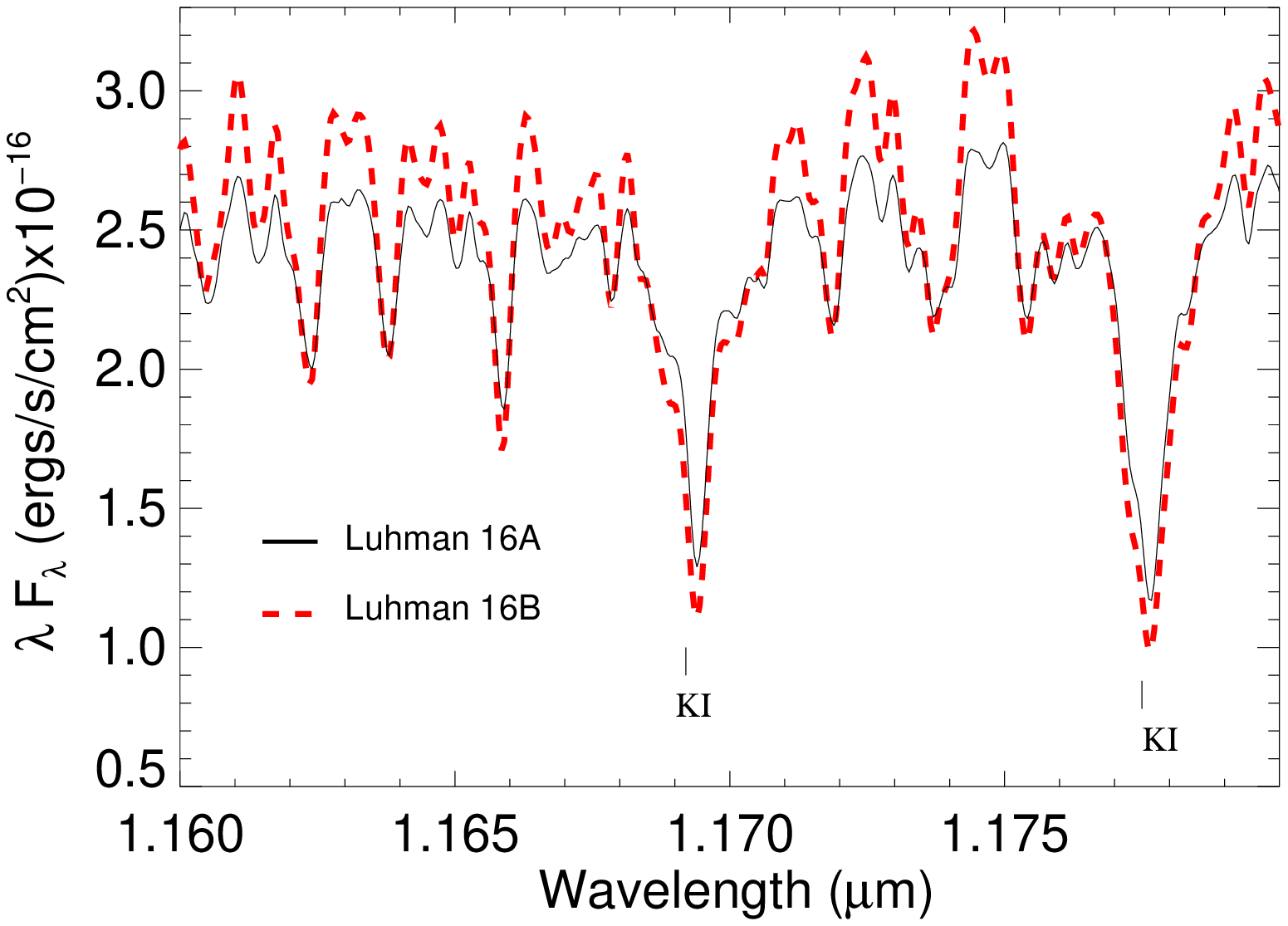}{(b)}&
\includegraphics[width=3.5in]{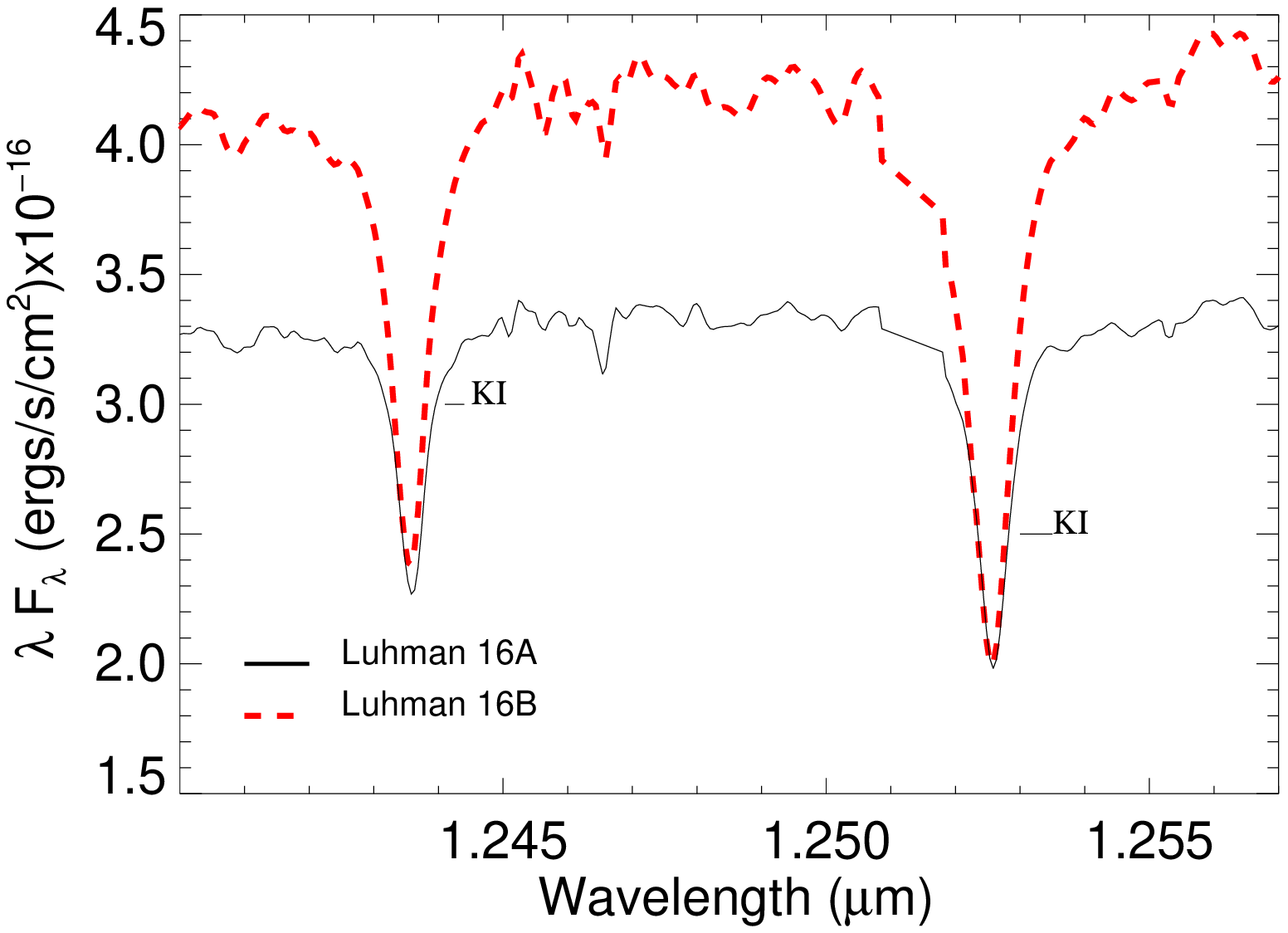}  \\
\end{array}$
\end{center}
\caption{{\bf (a)}  The normalized flux of Luhman 16A and Luhman 16B shown around the (1.168, 1.177 $\mu$m--Left) and (1.243, 1.254 $\mu$m--Right) K~I doublets.  Flux is normalized over the peak of the region shown and sources are offset from one another by 0.4 (left) and 0.3 (right).  Overplotted are DENIS0205 (black, long-dashed), an L7 (optical), and SDSS0423 (green, short-dashed), a  T0 (near-infrared), from \citet{McLean06}.  All sources are binned to the resolution of FIRE ($\lambda$/$\Delta\lambda$$\sim$ 8000).    {\bf (b)} The same regions as shown in the top panel except scaled to the distance of the system with no offset between components.}
\label{fig:K1normal}
\end{figure*}

\begin{figure*}[!ht]
\begin{center}$
\begin{array}{cc}
\includegraphics[width=3.5in]{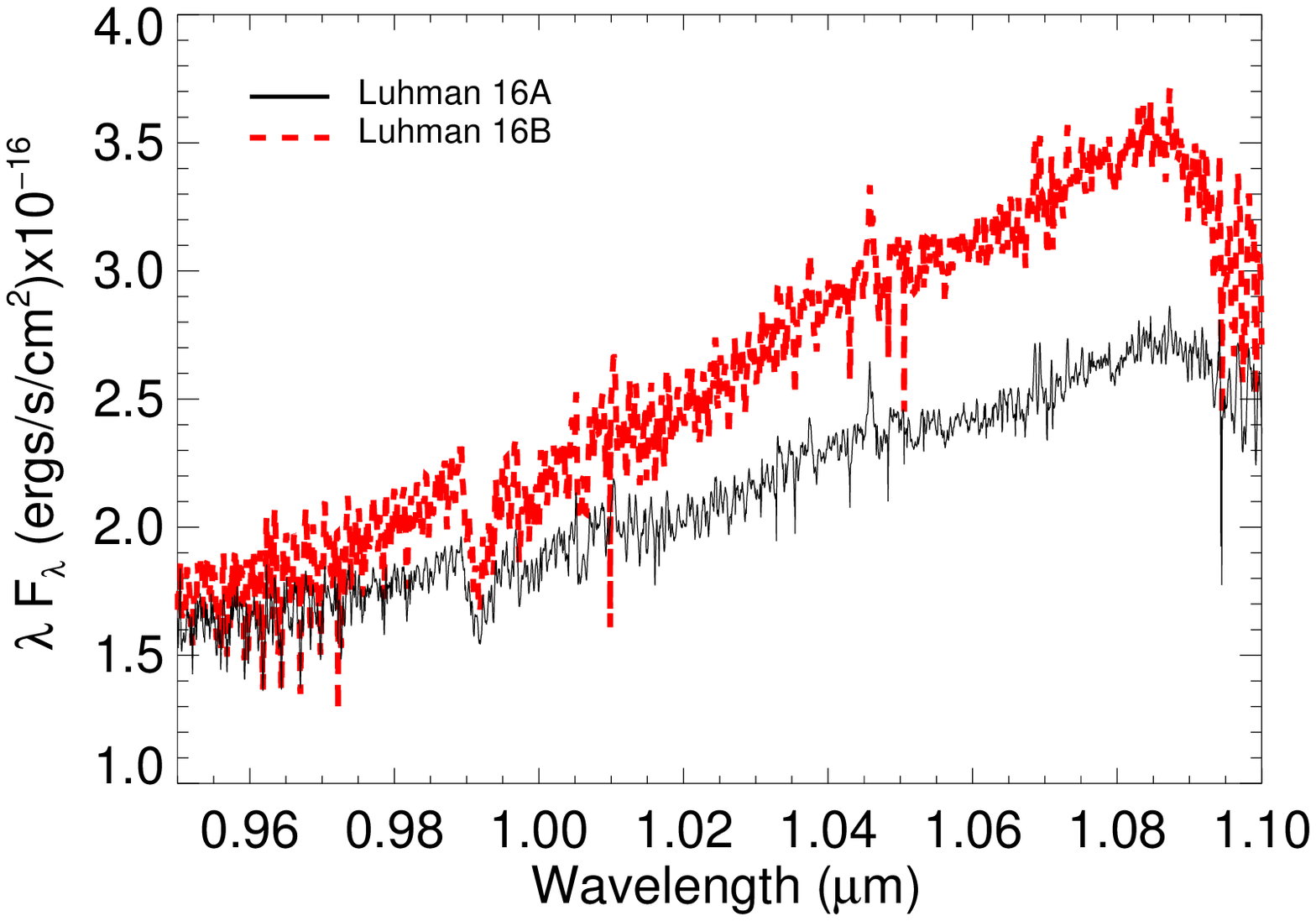} &
\includegraphics[width=3.5in]{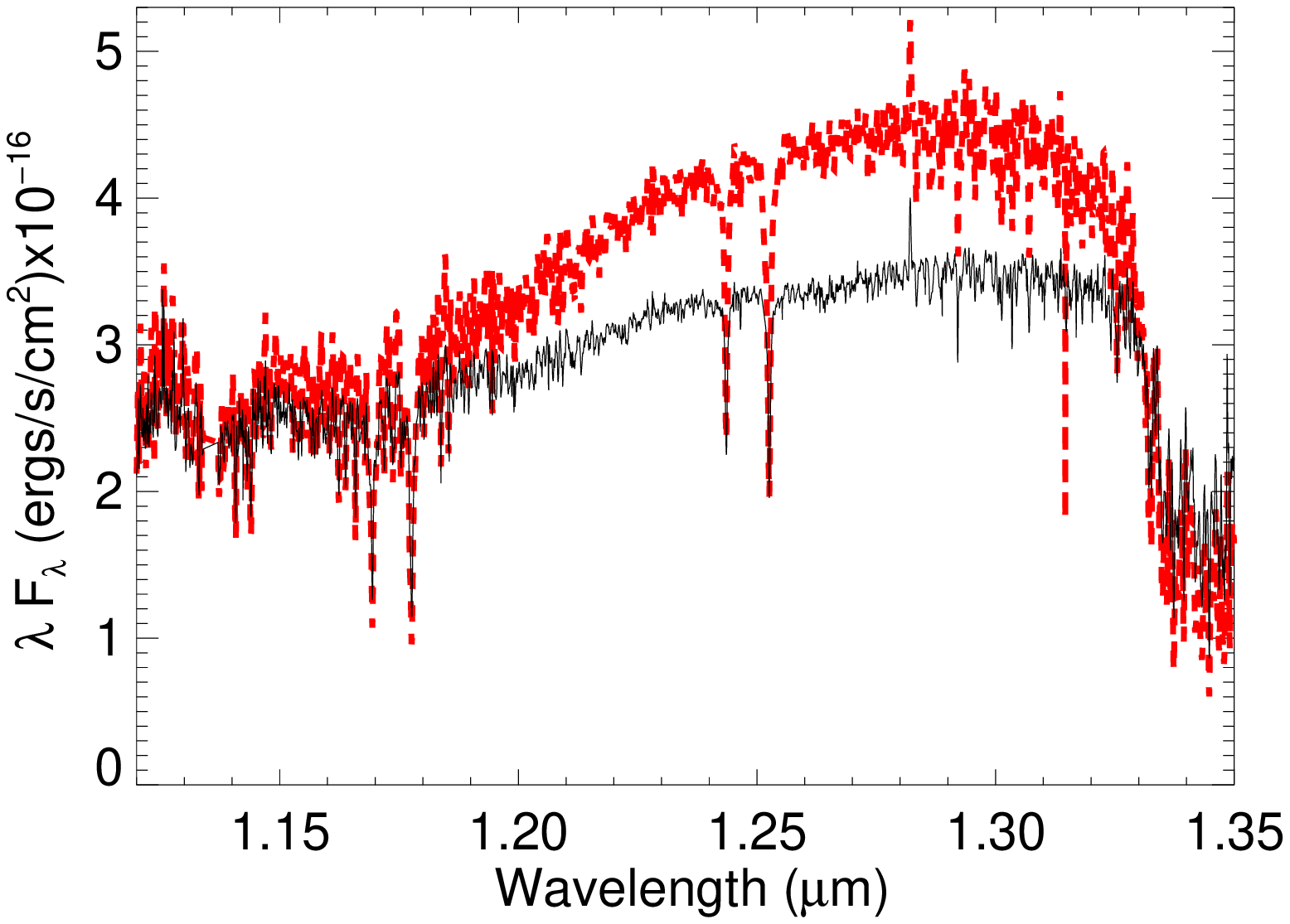} \\
\includegraphics[width=3.5in]{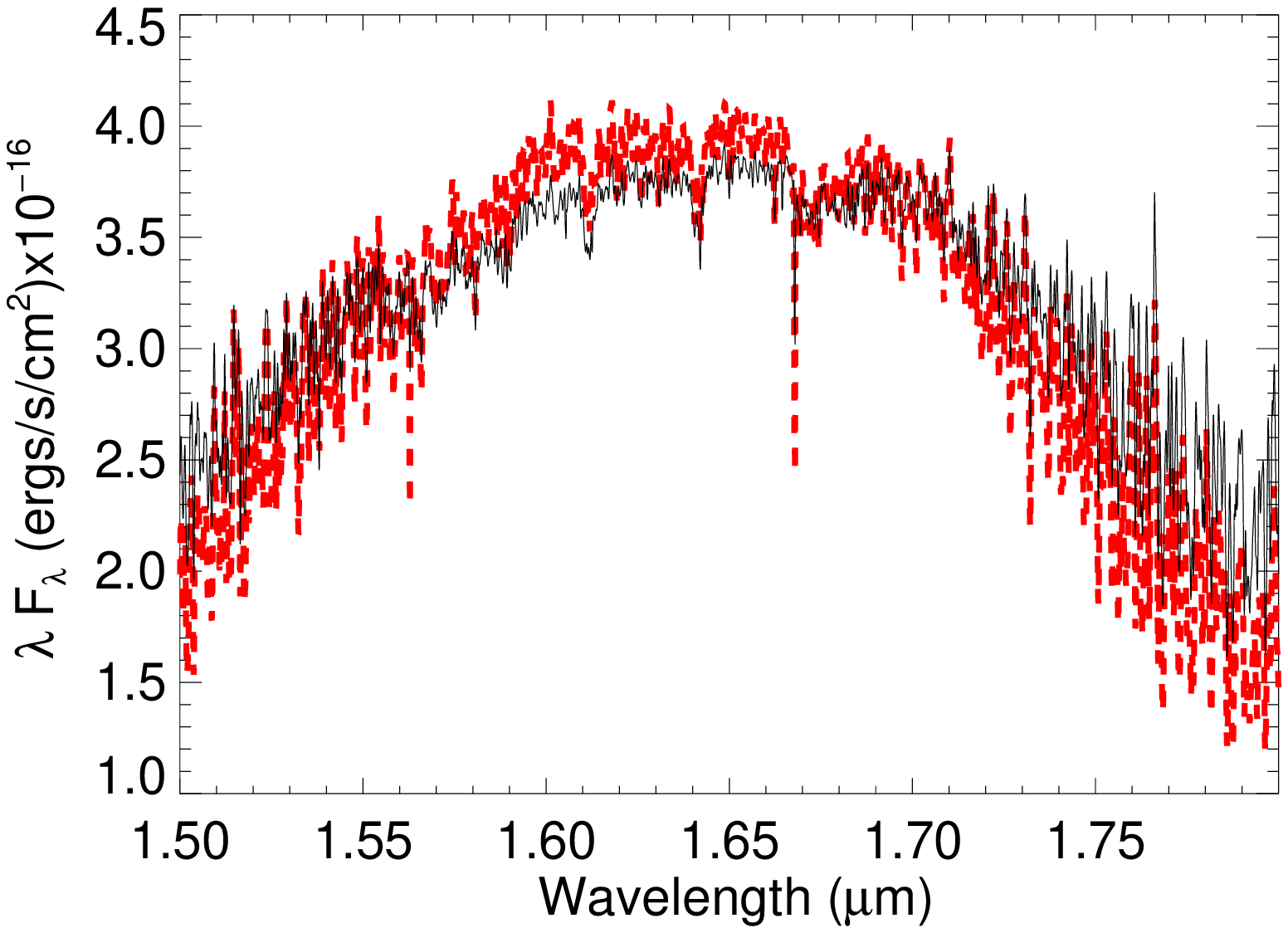}&
\includegraphics[width=3.5in]{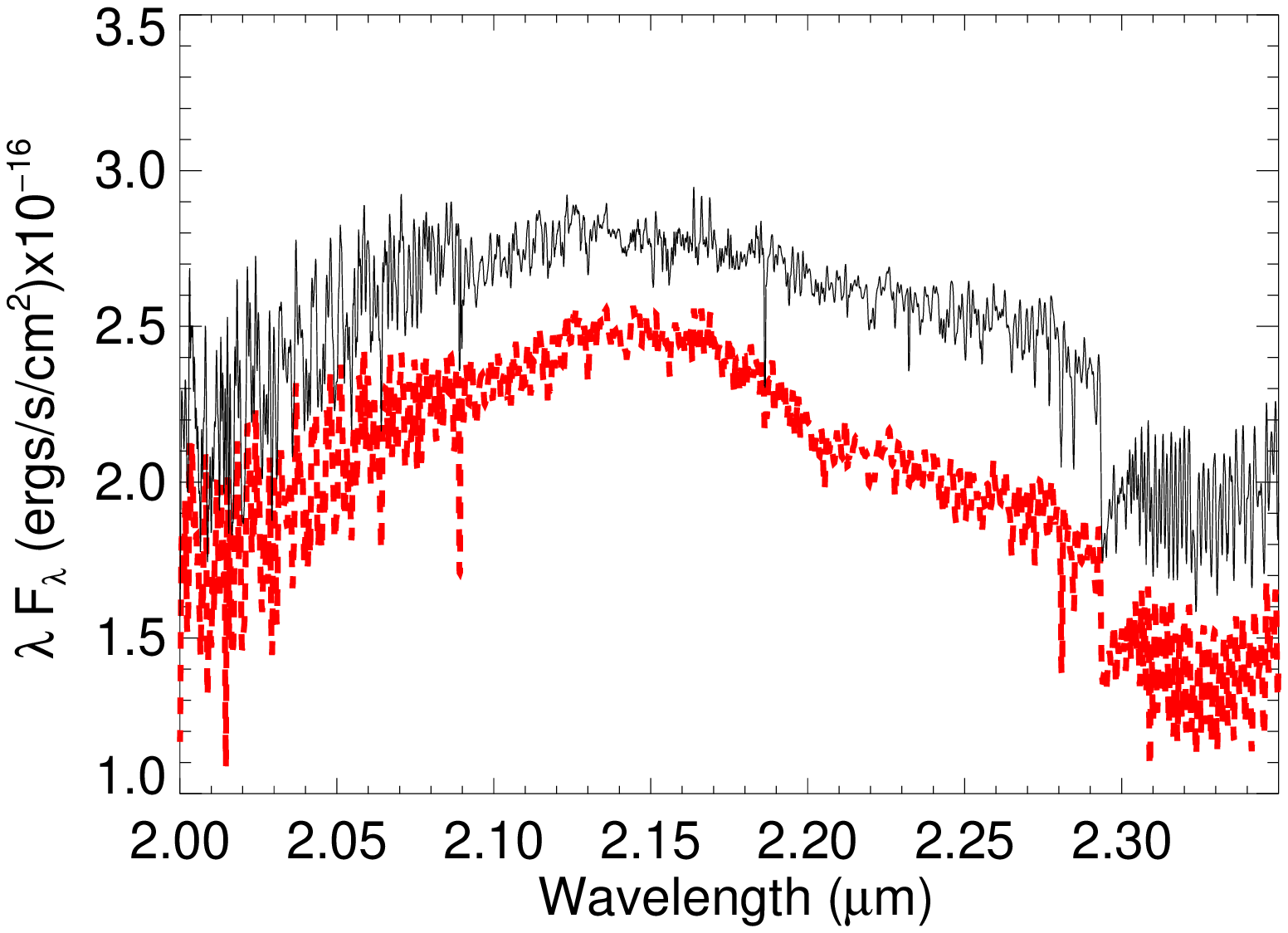}  \\
\end{array}$
\end{center}
\caption{The same regions shown in Figure~\ref{fig:spectra-bands} except scaled to the distance of the system with no offset between components. In $z$ and $J$ bands (top) the secondary is more luminous whereas this reverses by $K$ band (bottom right).}
\label{fig:reversal}
\end{figure*}

\section{DATA}
\subsection{FIRE Data}
On 28 March 2013 (UT) we used the 6.5m Baade Magellan telescope and the Folded-port InfraRed Echellette (FIRE; \citealt{Simcoe13}) spectrograph to obtain near-infrared spectra of each component in the Luhman 16AB system.  Observations were made under clear conditions with an average seeing of $\sim$0.5$\arcsec$ so we were able to easily resolve the two sources.  Each component was observed separately using the echellette mode and the 0.45$\arcsec$ slit  (resolution $\lambda$/$\Delta \lambda \sim$ 8000 at $J$ band) covering the full 0.8 - 2.5 $\micron$ band with a spatial resolution of 0.18$\arcsec$/pixel.  We first observed the A component using a 600s exposure, nodding 2$\arcsec$ in a North/South ABBA pattern to avoid contamination from the secondary.  We then moved to the B component and observed using an identical strategy.  Immediately after, we obtained two ThAr lamp spectra (21s and 63s) then observed the A0 V star HD 108196  (B=7.0, V=6.9)  fourteen times at 21s each in an ABBA pattern nodding by 2$\arcsec$.  At the end of the night we obtained dome flats and Xe flash lamps to construct a pixel-to-pixel response calibration.  Data were reduced using the FIREHOSE package which is based on the MASE and SpeX reduction tools (\citealt{Bochanski09}, \citealt{Cushing04}, \citealt{vacca03}). 

\subsection{MagE Data}
On 26 April 2013 (UT) we used the 6.5m Clay Magellan telescope and Magellan Echellette Spectrograph (MagE; \citealt{Marshall08}) to obtain optical spectra of each component in the Luhman 16AB system.  MagE is a cross--dispersed optical spectrograph, covering 3,000 to 10,000~\AA~ with a spatial resolution of 0.3$\arcsec$/pixel.  Our observations employed a $0.7^{\prime\prime}$ slit aligned at the parallactic angle (resolution $\lambda$/$\Delta \lambda \sim$ 4000 at $I$ band). Observations were made under clear conditions with an average seeing of $\sim$0.6$\arcsec$ so we were able to easily resolve the two sources with minimal contamination ($<$ 1\%).  A 1200s integration was obtained for Luhman 16A followed immediately by an identical observation of Luhman 16B and a 3s ThAr lamp spectrum for wavelength calibration.  The spectrophotometric standard GJ 318 was observed for flux calibration (180s). Ten Xe-flash lamp light spectra as well as dome flats were taken at the end of the evening for pixel response calibration.  The data were reduced using the MagE Spectral Extractor pipeline (MASE; \citealt{Bochanski09}) which incorporates flat fielding, sky subtraction and flux calibration IDL routines.

\section{SPECTRAL ANALYSIS}\label{spectralanalysis}
The combined MagE and FIRE spectral data are shown in  Figure~\ref{fig:fullnir} for both components of Luhman 16AB.  Each is scaled using the Mauna Kea Observatory (MKO) resolved photometry from \citet{Burgasser13} and the parallax from \citet{Boffin13}.  In general,  the overall shape of the components are comparable confirming the strong similarities in their effective temperatures (Luhman 16A, L7.5; Luhman 16B T0.5 see \citealt{Burgasser13}, \citealt{Luhman13}, \citealt{Kniazev13}).  The prominent differences distinguishing the spectral subtypes include differing slopes when moving from the optical into the near-infrared and stronger CH$_{4}$ absorption at 1.15$\micron$ and 2.2$\micron$ in Luhman 16B.   In the following Subsections, we break the spectra into narrow optical and infrared bandpasses and discuss signatures of temperature, gravity, and atmosphere conditions.

\subsection{Optical Data}
In Figure~\ref{fig:Mage} we present the MAGE spectra of both components highlighting the location of prominent molecular features. While the optical spectra of Luhman 16A and Luhman 16B have been presented in \citet{Luhman13} and \citet{Kniazev13} respectively, diagnostic features have yet to be explored in detail.  

The most notable optical feature is the clear detection of the 6708 \AA\  Li I absorption line in both Luhman 16A and Luhman 16B.  The core temperature required to ignite Lithium burning is lower than that required for Hydrogen burning.   In turn, this translates into a lower fusing mass limit ($\sim$ 0.065 M$_{\sun}$; \citealt{Rebolo92}, \citealt{Magazzu93}).  The interiors of lower mass stars and brown dwarfs are fully convective, therefore objects above this fusing mass limit will fully deplete their reservoir of Lithium (in $\sim$ $<<$ 1 Gyr; e.g. \citealt{Chabrier96} ) while those below it, will not.  Consequently, a detection of Lithium in ultracool dwarfs (T$_{eff}$ $<$ $\sim$ 2700; \citealt{Basri98})  implies a mass limt of $\sim$0.065 M$_{\odot}$ which can be translated into an age upper limit.  At the T$_{eff}$s discussed in Section ~\ref{Models}, we estimate an age upper limit of 3 Gyr for Luhman 16AB.

Interestingly, this is the first detection of Li I absorption in a T dwarf.  As discussed in \citet{Lodders06} and \citet{Lodders99}, at T$_{eff}s$ $<$ $\sim$1500 K,  lithium rapidly becomes bound in molecules such as LiCl and LiOH.  In support of this idea, \citet{Kirkpatrick08} present a detailed analysis of the optical spectra of L and T dwarfs and show that while the strength of the Li I 6708 \AA\ absorption line increases through $\sim$ L6, it rapidly weakens into the latest L dwarfs and is undetected in all T dwarfs at $>$ $\sim$ 4\AA\  (see also \citealt{Burgasser03}).  Additionally, \citet{King10} present a detailed spectral analysis of the (previously) closest T dwarf system, Epsilon Indi Bab (T1+T6), and find no evidence for lithium absorption at 6708 \AA\ . The \citet{King10} spectra were a factor of eight lower than the data in this paper (R $\sim$ 1000 for Epsilon Indi Ba as opposed to R$\sim$ 8000 for Luhman16A).  However as discussed in \citet{King10}, Epsilon Indi Ba (a T1) requires a lithium depletion of at least 1000 to remove the 6708 \AA\ absorption line.  This indicates that the strong detection reported for Luhman 16A in this work is significantly different than that of the previously best studied early T dwarf.   

We report the Li I absorption equivalent widths (EW)\footnote{All equivalent widths are measured with respect to a pseudo-continuum therefore should be considered pseudo-equivalent widths throughout} for both components in Table ~\ref{Table:EQ_OPT}.  Luhman 16A, an L8.5, has a Li I absorption EW of 8.0$\pm$0.4 \AA\ consistent with the median Li I EW for L7-L8 dwarfs with measurable detections in \citet{Kirkpatrick08} ($\sim$ 40\% of their L8 sample had median EW of  $\sim$ 9-10 \AA).   Luhman 16B has appreciably lower absorption (EW $\sim$ 3.8 $\pm$0.4) but the line is clearly detected in the inset of Figure ~\ref{fig:Mage}. We have also marked the expected position of the 6562.8 \AA\  H $\alpha$ line in the inset of Figure ~\ref{fig:Mage} however there is very little flux in this region and we find only an upper limit for emmision or absorption of  1.5 \AA\ .

Figure~\ref{fig:Mage} also highlights the presence of K I, Rb I and Cs I lines as well as the broadband CrH+FeH feature.  The Cs I lines have a relatively weak dependence on gravity and have been used as a spectral index to estimate T$_{\mathrm eff}$ (see e.g. \citealt{Lodders99}, \citealt{Burgasser03}, \citealt{Kirkpatrick99}).   In particular, the 8521\,\AA\  and 8943\,\AA\  Cs I lines are found to increase in strength through the L dwarfs and peak at optical spectral types of T2 before declining through late-type T's (\citealt{Kirkpatrick99}, \citealt{Burgasser03}).  Similarly, the 7800\,\AA\  and 7948\,\AA\  Rb I lines are found to strengthen through the L dwarfs.  However they lie very close to the core of a strong pressure-broadened K I doublet in the optical data of T dwarfs so their trends in that temperature regime are more difficult to quantify.  

We find that, as expected, the T0.5 secondary Luhman 16B, has stronger (or comparable) Cs I and Rb I than the L7.5 primary Luhman 16A.   We report equivalent widths for each line in Table ~\ref{Table:EQ_OPT} measured in a similar manner to that described in \citet{Burgasser03}.   We find our values are comparable to those for late L dwarfs and early T dwarfs in \citet{Kirkpatrick99} and \citet{Burgasser03}.  We note that 
the MagE CCD is known to show fringing in the red region of the spectrum starting at 7000\,\AA\ and can reach peak amplitudes of up to 10\%. Incandescent lamp flats were used to correct for this effect, however there appears to be residual fringing long ward of 8500\,\AA\  that may contribute to a poor sampling of the Cs I lines.  Figure~\ref{fig:Mage} also highlights the expected location of the Na I doublet (8183 - 8195 \AA).  While fringing and telluric features do impact this area of the spectrum, we find no evidence for Na I absorption in either source at $>$ 0.5 \AA\ (see also the near infrared analysis in Section 3.5).

\subsection{Z band}
Figure~\ref{fig:spectra-bands}a shows the 0.95 - 1.10 $\mu$m FIRE $z$ band data with features of FeH, CH$_{4}$ and H$_{2}$O highlighted.  The most prominent is the Wing-Ford band (\citealt{Wing69}) of FeH starting at 0.9896 $\mu$m.  FeH is known to be an important opacity source in the atmospheres of brown dwarfs (\citealt{Cushing03}).   The Wing-Ford band specifically is very strong in M dwarfs then declines through mid-L's as FeH condenses out of the atmosphere forming a cloud layer below the detectable photosphere.  However it re-appears in early T dwarfs as a decreasing T$_{\mathrm eff}$ disrupts cloud layers leading to holes that allow the observation of deeper/hotter layers (e.g. \citealt{Burgasser02}).  As discussed in Section ~\ref{section:clouds}, Luhman 16A may be cloudy, but Luhman 16B is thought to have an atmosphere with rapidly evolving cloud patterns (\citealt{Gillon13}, \citealt{Biller13}, \citealt{Crossfield14}, \citealt{Burgasser14}).  The presence of comparably strong FeH in both components implies that the underlying photospheres of Luhman 16A and Luhman 16B are similar despite the fact that only the secondary shows strong weather related phenomenon.

\subsection{H band}
In Figure~\ref{fig:spectra-bands}c  we show the full resolution 1.45 - 1.80 $\mu$m $H$ band data with molecular features of FeH and CH$_{4}$ highlighted.  Gravity impacts the shape of the $H$ band. At younger ages (hence lower gravities),  collisionally induced H$_{2}$ absorption in $K$ band is lessened, and this sculpts the longer wavelength side of the $H$ band into a triangular shape. This is a known feature of Pleiades ($\sim$ 120 Myr) and younger late-type M and early-mid L dwarfs (see \citealt{Lucas01}, \citealt{Allers07}, \citealt{Rice10}, \citealt{Kirkpatrick06}, \citealt{Faherty13, Faherty13a}, \citealt{Gizis12}, \citealt{Bihain10}).   The $H$ band shape for Luhman 16A and 16B are similar, and show no sign of a lower surface gravity.  This coincides with our analysis of the alkali lines (see Section 3.5) and implies the system is likely older than 120 Myr.

The FeH features at 1.60 $\mu$m, and 1.63 $\mu$m are comparable in each component as is the 1.67 $\mu$m CH$_{4}$ feature. Both are thought to strengthen with decreasing T$_{\mathrm eff}$, thus indicating that the temperatures of Luhman 16A and Luhman 16B are very similar.   

\subsection{K band}
In Figure~\ref{fig:spectra-bands}d we show the 2.0 - 2.35 $\mu$m $K$ band data with molecular features of CH$_{4}$ and CO highlighted.   Comparing the two components, the $K$ band shape shows the strongest difference between Luhman 16A and Luhman 16B as the 2.20 $\mu$m band head of CH$_{4}$ is much stronger in the secondary.  This is the clearest indication of the later spectral type and expected lower temperature of Luhman 16B.  

In general, the $K$ band offers a lever for gauging metallicity and gravity effects as it is suppressed with decreasing metallicity and/or increasing gravity and enhanced for lower surface gravity and/or higher metallicity (e.g. \citealt{Burgasser06}).  Examining all spectral features (including $K$ band) that are indicative of metallicity and gravity effects as a whole, we find that the components do not deviate significantly from the expectation of a field aged L or T dwarf. Furthermore, we conclude that the components show temperature differences, but nothing sufficiently striking as to indicate that either gravity or metallicity are at all different in the two components.

\subsection{J band}\label{Jband}
Figure~\ref{fig:spectra-bands}b shows the 1.12 - 1.35 $\mu$m normalized $J$ band data with molecular features of FeH, CH$_{4}$, and H$_{2}$O as well as the alkali doublets of K~I labeled.  \citet{Burgasser13} discuss the alkali spectral features in low-resolution FIRE and SpeX prism data citing strong signatures of K~I, and hints of Na I in each component.  As shown in Figure~\ref{fig:Na I}, we find no trace of the Na I  doublet (1.138,1.141) $\mu$m in either. However, the K~I doublets at (1.168, 1.177) $\mu$m and (1.243, 1.254) $\mu$m are indeed very strong. We report equivalent widths for each line in Table ~\ref{Table:EQ_IR}. 

For brown dwarfs, the most prominent trends found in studies of the alkali lines are linked to: (1) a temperature dependence and (2) a gravity dependence.  In the case of (1), the strength of the 1.17 $\mu$m and 1.25 $\mu$m K~I doublets show two peaks at $\sim$ L4 and T3 with mid to late- L dwarfs falling in the trough between (see Figure~\ref{fig:mclean}; and \citealt{McLean03,McLean06}, \citealt{Burgasser02}, \citealt{McGovern04}).   This effect is consistent with the idea that we probe much greater depths in cool T dwarfs and the line-width and depth of alkali lines is related to atmospheric chemistry (altered by a changing T$_{\mathrm eff}$).  In the case of (2), younger objects  have not contracted to their final radii so they have a lower surface gravity hence lower atmospheric pressure. The consequences of which are less pressure broadening and narrower alkali lines  (e.g. \citealt{McGovern04}, \citealt{Allers07}, \citealt{Kirkpatrick06}, \citealt{Cruz09}, \citealt{Rice10, Rice11}, \citealt{Faherty13}). 

To test gravity and/or temperature indications from the strength and depth of the alkali lines, we compare the spectral region around each K~I doublet to a well-studied comparable subtype (e.g. probe of T$_{\mathrm eff}$) source and we compare equivalent widths with a sample of late-type M, L, and T dwarfs.  Figure~\ref{fig:mclean} shows the latter, comparing K~I line equivalent widths of 53 ultracool dwarfs from the \citet{McLean03} low-resolution (R$\sim$ 2000) NIRSPEC dataset to our measurements for both components.  We binned our higher resolution data to that of the \citet{McLean03} sample and followed their prescription for determining the continuum level and line-width range.  Uncertainties in equivalent width were calculated via the method outlined by \citet{Looper08} using measurements of multiple noise spikes.  The uncertainty in spectral type for most sources examined is $\pm$0.5 subtype, therefore we conclude that both components fall within the trends set by the large ultra cool dwarf sample.  Interestingly, Luhman 16A tends toward weaker lines and Luhman 16B tends toward stronger lines.  Given their similar T$_{eff}$s and the coeval nature of the system, this is likely a signpost of atmosphere conditions (i.e clouds).  

In Figure ~\ref{fig:K1normal}a, we directly compare each spectrum to that of DENIS-P J0205.4-1159 (DENIS0205), an L7 (optical), and SDSSp J042348.57-041403.5 (SDSS0423), a  T0 (near-infrared), from the \citet{McLean06} sample. We note that DENIS0205 was the closest in spectral subtype to Luhman 16A however it is a confirmed binary and potential triple system  (L5, L8, T0, \citealt{Bouy05}).  Unresolved binarity will impact the interpretation of the alkali line trends as the components (and the effects on their lines) are blended.  In this case, the inferred late-type components of DENIS-0205 dominate its alkali line trends. Since they are close in nature to Luhman 16A and Luhman 16B, a comparison should be valid.  

Both DENIS0205 and SDSS0423 were observed with NIRSPEC on Keck at a resolution of $\lambda$/$\Delta\lambda$$\sim$40,000~\AA~ so we had to first bin them down to the FIRE echelle resolution of $\lambda$/$\Delta\lambda$$\sim$8,000 using the IDL ``smooth" function.  Figure~\ref{fig:K1normal}a shows a zoomed in view of the (1.168, 1.177) $\mu$m and (1.243, 1.254) $\mu$m K~I line doublets normalized over the peak of the displayed region.  Using this normalization approach, it appears that Luhman 16A has narrower and weaker K~I doublets than the standards and the B component.  Conversely, Luhman 16B matches well to the standard for the (1.168, 1.177) $\mu$m K~I doublet but shows deeper absorption for the  (1.243, 1.254) $\mu$m K~I doublet.

In Figure~\ref{fig:K1normal}b we show the spectral regions around the K~I doublets scaled to the distance of the system.  Using this comparison removes the arbitrary  normalization applied to the components that can skew analyzing the line profiles.   We find that the differences between components is not broader/weaker K~I line features (hence a gravity indication),  but rather brighter/fainter continuum (see also \citealt{Burgasser13a}). For the (1.243, 1.254) K~I doublet , Luhman 16B is more luminous than Luhman 16A.  We discuss these differences in terms of potential cloud variations in Section ~\ref{section:clouds} below.

\section{CLOUDS IN THE COMPONENTS OF LUHMAN 16AB }\label{section:clouds}
According to \citet{Gillon13}, the Luhman 16 system shows strong photometric variability across its quasi-periodic (P=4.87 $\pm$0.01h) light curve (see also \citealt{Biller13}, \citealt{Burgasser14}).  The peak to peak amplitude change of up to 11\% at 1$\micron$ is attributed to weather patterns with rapidly changing cloud structures in only the secondary, Luhman 16B (see \citealt{Crossfield14}). In theory, both Luhman 16A and Luhman 16B are in the prime spectral type range for rapid cloud-clearing.   As suggested in both \citet{Burgasser13} and \citet{Gillon13} the Luhman 16AB system must straddle the thin boundary in temperature/mass where cloud clearing occurs.

\begin{figure*}[!ht]
\begin{center}$
\begin{array}{cc}
\includegraphics[width=3.5in]{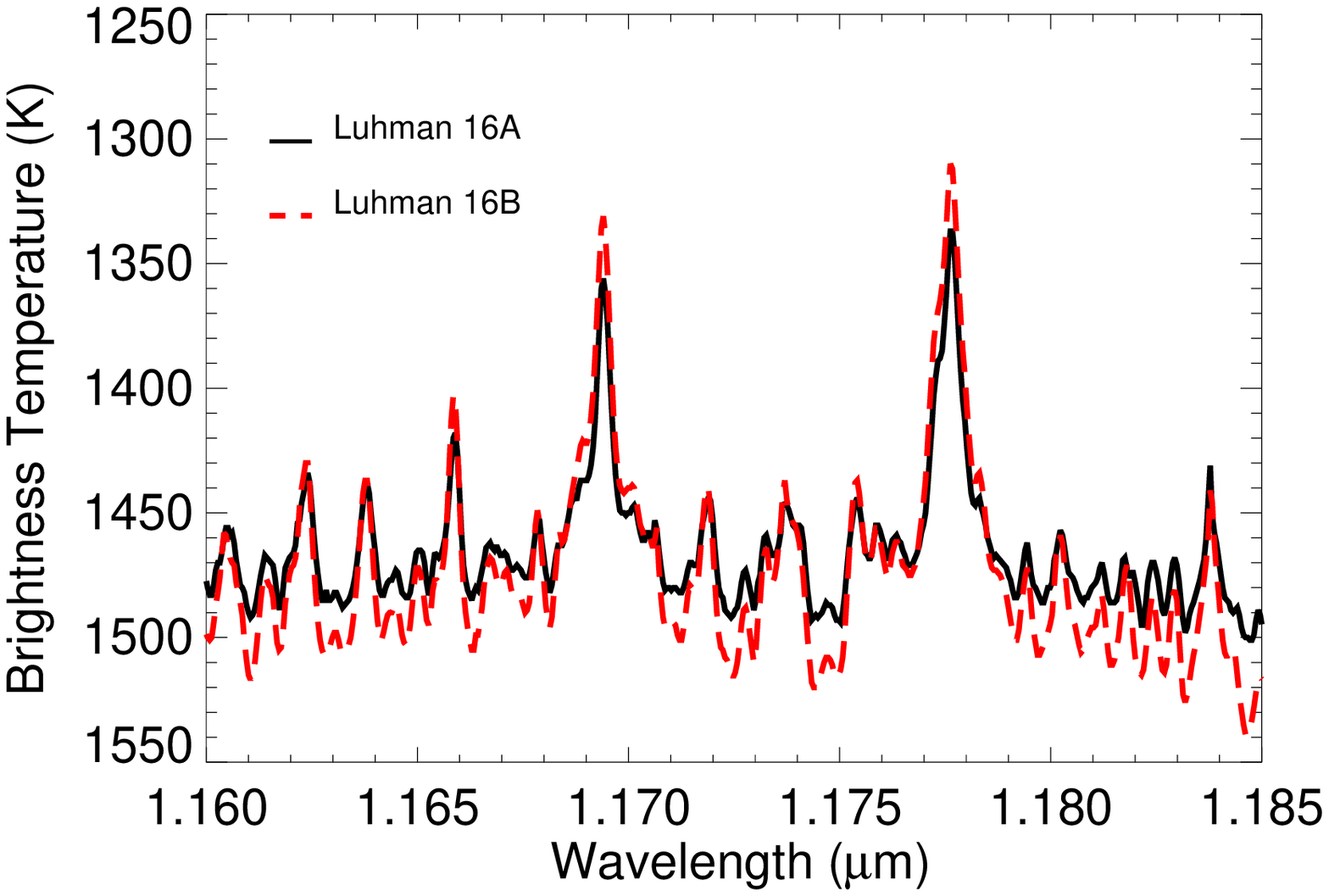}&
\includegraphics[width=3.5in]{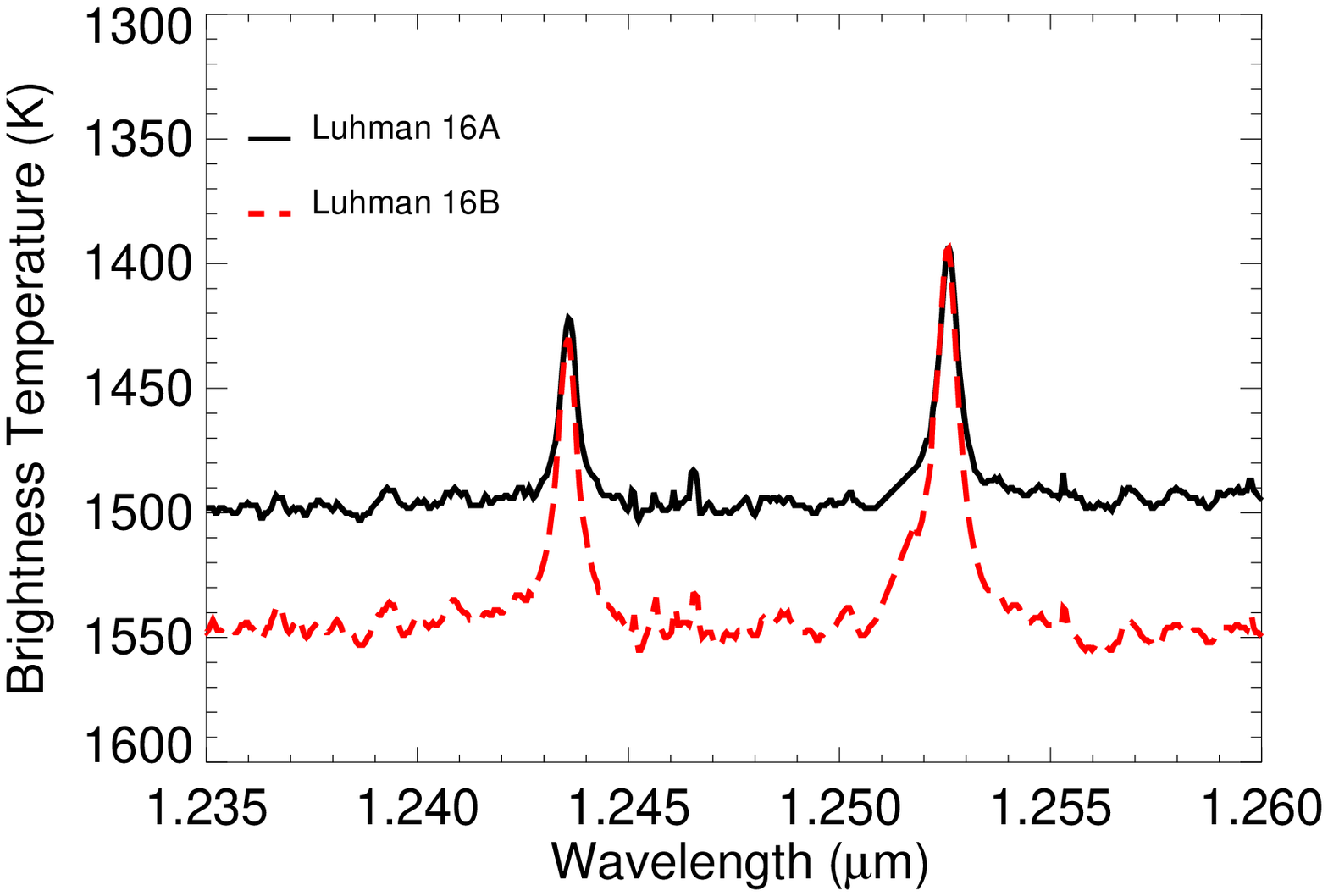}  \\

\end{array}$
\end{center}
\caption{The same regions as shown in Figure~\ref{fig:K1normal} except flux has been converted into a brightness temperature by transforming the observed flux densities to surface densities using the absolute J magnitudes reported in \citet{Burgasser13} and a radii of 0.90~R$_{Jup}$ (based on the evolutionary models of \citealt{Burrows01}).  At each wavelength, we determine the temperature (T) for which a corresponding blackbody distribution, $\pi{B_{\lambda}}$(T), produces the same intensity. }
\label{fig:K1absolute}
\end{figure*}

 \begin{figure}[ht!]
\resizebox{1.0\hsize}{!}{\includegraphics[clip=true]{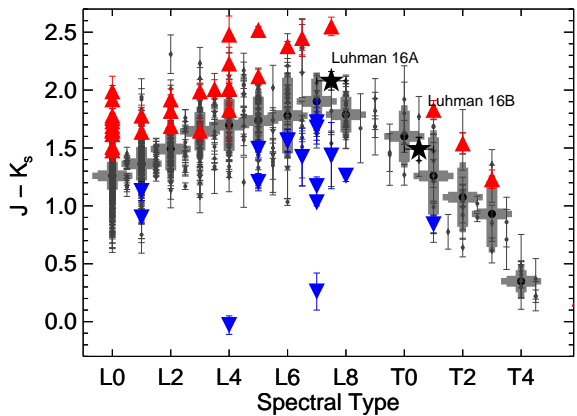}}
\caption{Spectral Type vs. 2MASS (J-K$_{s}$) color for L0-T4 dwarfs.  Median values and their spread from \citet{Faherty13} (L dwarfs) and this work (T dwarfs) are shown as grey boxes.  Individual sources collected from dwarfarchives.org are over plotted with uncertainties as are low-gravity, dusty, and unusually red sources (red upward facing triangles) and subdwarfs, unusually blue, and peculiar sources (blue downward facing triangles). All red and blue sources were compiled from \citet{Kirkpatrick10}, \citet{Faherty09}.  Component photometry for Luhman 16A and Luhman 16B from \citet{Burgasser13} are converted to 2MASS magnitudes using the \citealt{Stephens04} relations. They are over plotted as black five-point stars.  } 
\label{fig:jmk}
\end{figure}

 \begin{figure*}[!ht]
\begin{center}$
\begin{array}{cc}
\includegraphics[width=3.5in]{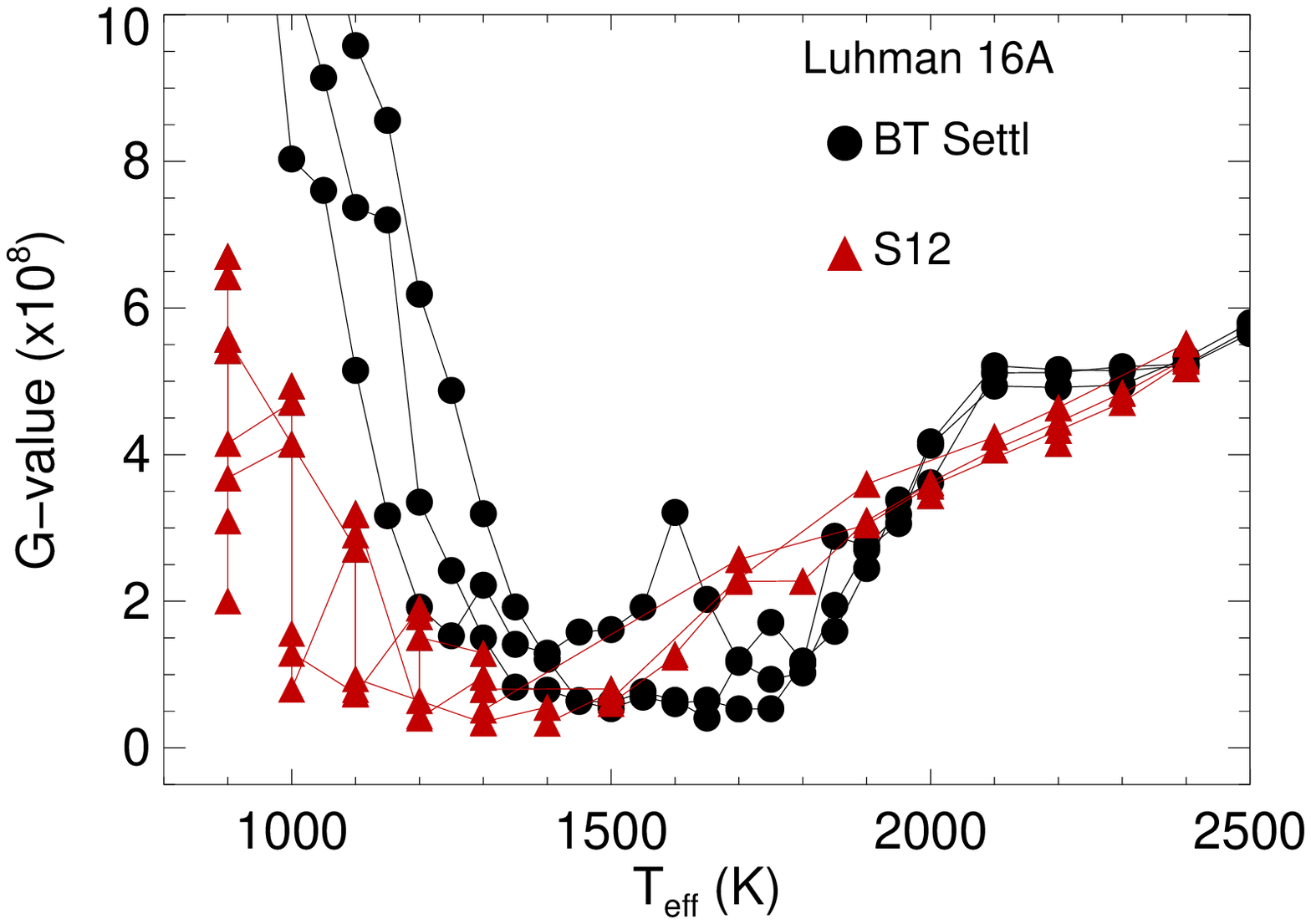} &
\includegraphics[width=3.5in]{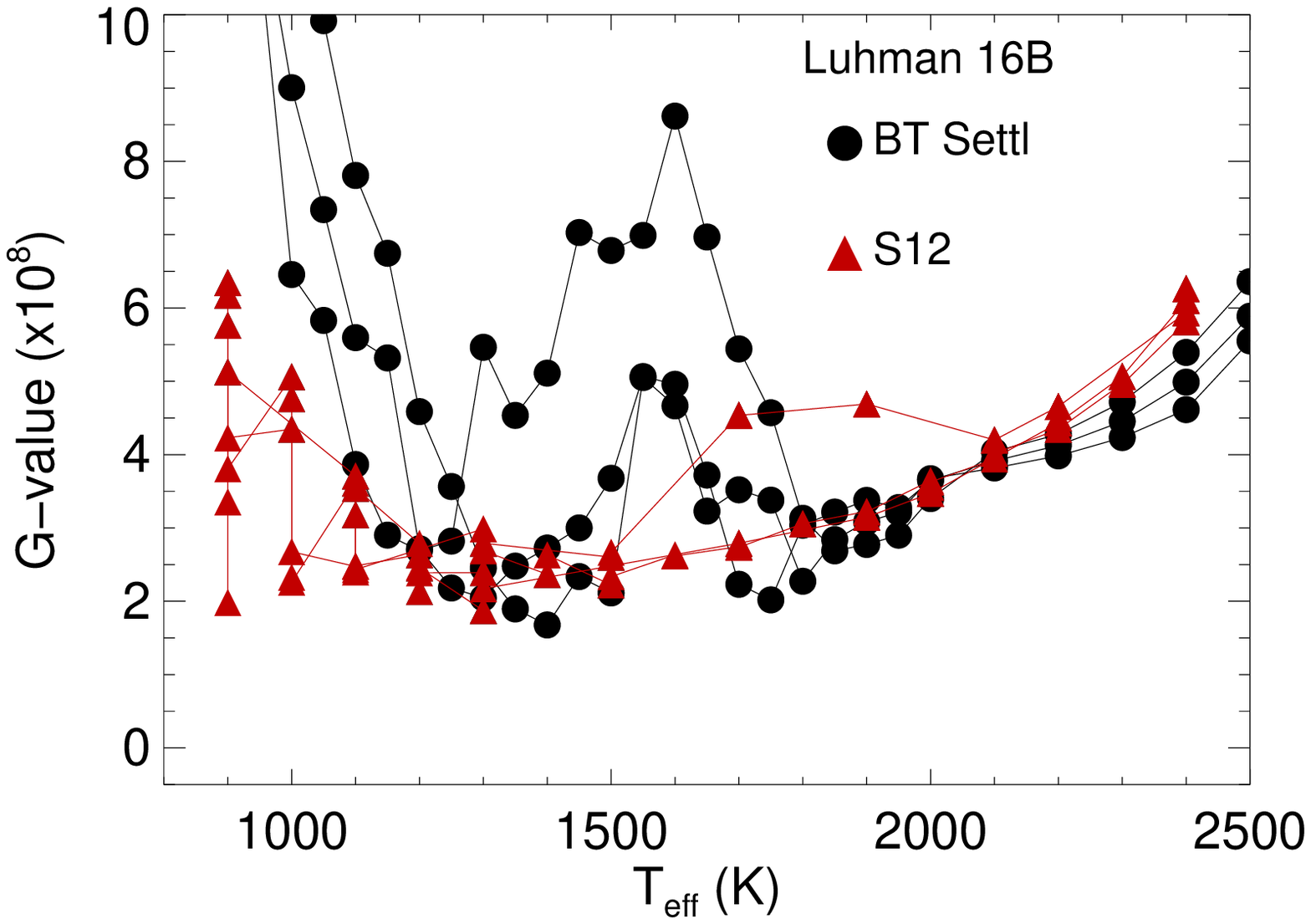} \\
\end{array}$
\end{center}
\caption{The model T$_{\mathrm eff}$ versus G-value (goodness of fit statistic) for the near-infrared data on Luhman 16A (left panel) and Luhman 16B (right panel) when fit to synthetic spectra.  In this work we compare to the BT Settl (black filled circles), and  S12 models (red upward facing triangles for  S12 models including clouds, and blue downward facing triangles  for  S12 models excluding clouds).  In the case of both models the range of fits shown at each T$_{\mathrm eff}$ also include ranging gravities (BT Settl \&  S12)  and equilibrium chemistry (S12).  Metallicity is assumed to be solar.  Marked by a vertical line on each panel is the location of the minimum G-value or best fit for each model.  Parameters are reported in Table ~\ref{Table:Models}.}
\label{fig:Gvalues}
\end{figure*}

\subsection{Luhman 16AB as a Flux reversal Binary}
Binaries that span the L/T boundary demonstrate a flux reversal whereby the cooler secondary is brighter in $z$ and $J$ bands than the warmer primary (e.g. \citealt{Gizis03}, {\citealt{Looper08}, \citealt{Liu06}, \citealt{Burgasser06a}).  The mechanism that causes this flux reversal and the corresponding $``J"$ band bump in brown dwarf evolutionary diagrams (where early T dwarfs are up to 0.5 mag more luminous at $J$ than slightly warmer sources -- \citealt{Tinney03}, \citealt{Vrba04}, \citealt{Dupuy12}, \citealt{Faherty12}) is predicted to be rapid cloud clearing as objects transition from cloudy L dwarfs to relatively clear T dwarfs (e.g.  \citealt{Ackerman01},\citealt{Burgasser02}). As discussed in \citet{Burgasser13}, noted in \citet{Boffin13} and shown in Figure~\ref{fig:reversal}, the Luhman 16AB system follows this trend with the secondary being 0.31$\pm$0.05 mag brighter at $J$ band and visually brighter in $z$ band.    

Focusing on the $J$ band region where the flux reversal is largest, we investigate the continuum regions around the K~I lines shown in Figure~\ref{fig:K1normal}b. We find a significant difference in flux between regions dominated by condensate grain opacity (the continuum around 1.25 $\mu$m) and regions dominated by molecular gas opacity (the continuum around 1.17 $\mu$m--\citealt{Ackerman01}).   Without knowing if Luhman 16A, Luhman 16B or both were varying at the time the data were taken, we cautiously view their flux differences in terms of a temperature gradient. To do this, we transform the observed flux densities to surface densities using the absolute J magnitudes reported in \citet{Burgasser13} and a radii of 0.90~R$_{Jup}$ (based on the evolutionary models of \citealt{Burrows01}).  At each wavelength, we determine the temperature (T) for which a corresponding blackbody distribution, $\pi{B_{\lambda}}$(T), produces the same intensity.  Figure~\ref{fig:K1absolute} shows the results for the area around both sets of K~I alkali lines.  At 1.25 $\mu$m, we find that Luhman 16B is $\sim$ 50$\pm$10 K warmer than Luhman 16A and at 1.17 $\mu$m we find Luhman 16B is $\sim$ 10$\pm$2 K warmer.     Uncertainties are conservatively estimated at 20\% given that they are dominated by uncertainties in the distance, photometry, and radii for both components (radii may vary at 0.90$\pm$0.15~R$_{Jup}$ and the system distance is 2.02$\pm$0.019pc).   We conclude that the brightness temperature difference between components at 1.17 $\mu$m is dominated by a T$_{\mathrm eff}$ distinction while at 1.25 $\mu$m it is the signature of cloud structure variations.  Luhman 16B may be the cooler source, but at 1.25 $\mu$m it is warmer because either a thinner cloud layer is present or atmospheric holes are allowing flux to emerge from warmer layers.  

\subsection{Near Infrared Photometry Indicators of Clouds Among Components}
The potential atmospheric conditions of Luhman 16A and Luhman 16B can also be discussed in the context of broadband near-infrared photometric properties.  In Figure~\ref{fig:jmk} we show the spectral type versus 2MASS (J-K$_{s}$) color diagnostic for the field population with component photometry for Luhman 16AB highlighted (photometry from \citealt{Burgasser13} converted to 2MASS using the \citealt{Stephens04} relations).   Both components are redward of the median for their given spectral subtypes and Luhman 16A is more than 1$\sigma$ from equivalent types.  In general, the reddest individual L dwarfs are those classified as having a low-surface gravity and suspected as harboring thick photospheric clouds (red triangles in Figure~\ref{fig:jmk} --e.g. \citealt{Cruz09}, \citealt{Faherty09, Faherty13,Faherty13a}, \citealt{Kirkpatrick10}).  The bluest L dwarfs are low-metallicity, potentially old sources (blue triangles in Figure~\ref{fig:jmk}--e.g. \citealt{Burgasser08}, \citealt{Cruz07}, \citealt{Burgasser04}, \citealt{Cushing09}, \citealt{Faherty09}, \citealt{Kirkpatrick10}).  Interestingly, as shown by spectral monitoring in \citet{Apai13} and \citet{Buenzli12}, brightness variations in L/T transition brown dwarfs occur without strong color changes since they find that the entire $J$ and $H$ band continuum brightens and dims.  By simultaneously changing cloud structure (thin to thick) and temperature (up to 300 K differences),  \citet{Apai13} find they can model the amplitude variations seen in L/T transition objects. \citet{Burgasser14} recently presented a resolved near-infrared spectroscopic monitoring campaign of the Luhman 16AB system and found that while the primary did not vary, the secondary did and a combination of achromatic (brightness) and chromatic (color) variability could explain its spectral variations.   Consistent with the \citet{Apai13} result, the color variation in Luhman 16B was small.  Consequently, we can infer that the redder color in Luhman 16A indicates thicker clouds hence a cooler brightness temperature at 1.25 $\mu$m.

\section{MODEL FITTING}\label{Models}
As Luhman 16AB are now the closest brown dwarfs known, their spectra will logically become an anchor for testing and advancing theoretical models.  As such we report the parameters from and discuss the quality of fits to the latest atmosphere model spectra.  We test synthetic data readily available from the BT Settl models (\citealt{Allard12}) using the \citet{Caffau11} solar abundances (referred to as CIFIST2011) as well as those generated (private communication) from the \citet{Saumon12} models (hereafter S12 models).  We used the distance scaled spectra shown in Figure~\ref{fig:fullnir} compared to a grid of model spectra with parameters ranging from T$_{\mathrm eff}$ (900 K -2400 K) and logg (4.5-5.5) at solar metallicity for BT Settl and T$_{\mathrm eff}$ (900 K -2200 K), logg (4.5-5.5) and chemical equilibrium (in or out) for a cloudy photosphere (fsed=2) from  S12.  We applied the model fitting technique described in detail in \citet{Cushing08} which uses a goodness of fit statistic, G$_{k}$, to determine the best fit model spectra:

\begin{equation}
G_{k}=\sum_{i=1}^{n}\left(\frac{f_{i}-C_{k}F_{k,i}}{\sigma_{i}}\right)^{2}
\end{equation}

\noindent where $f_{i}$ and F$_{k,i}$ are the flux densities of the data and model $k$, respectively; $\sigma_{i}$ are the errors in the observed flux densities; and C$_{k}$ is determined by minimizing G$_{k}$ and given by

\begin{equation}
C_{k}=\frac{\sum f_{i}F_{k,i}/\sigma_{i}^2}{\sum F_{k,i}^{2}/\sigma_{i}^2}
\end{equation}

The value for C$_{k}$ is the multiplicative constant required to match the synthetic spectra flux to observed data and is equal to (R/d)$^{2}$, where $R$ is the objects radius, and $d$ is the objects distance.  

\begin{figure*}[ht!]
\resizebox{1.0\hsize}{!}{\includegraphics[clip=true]{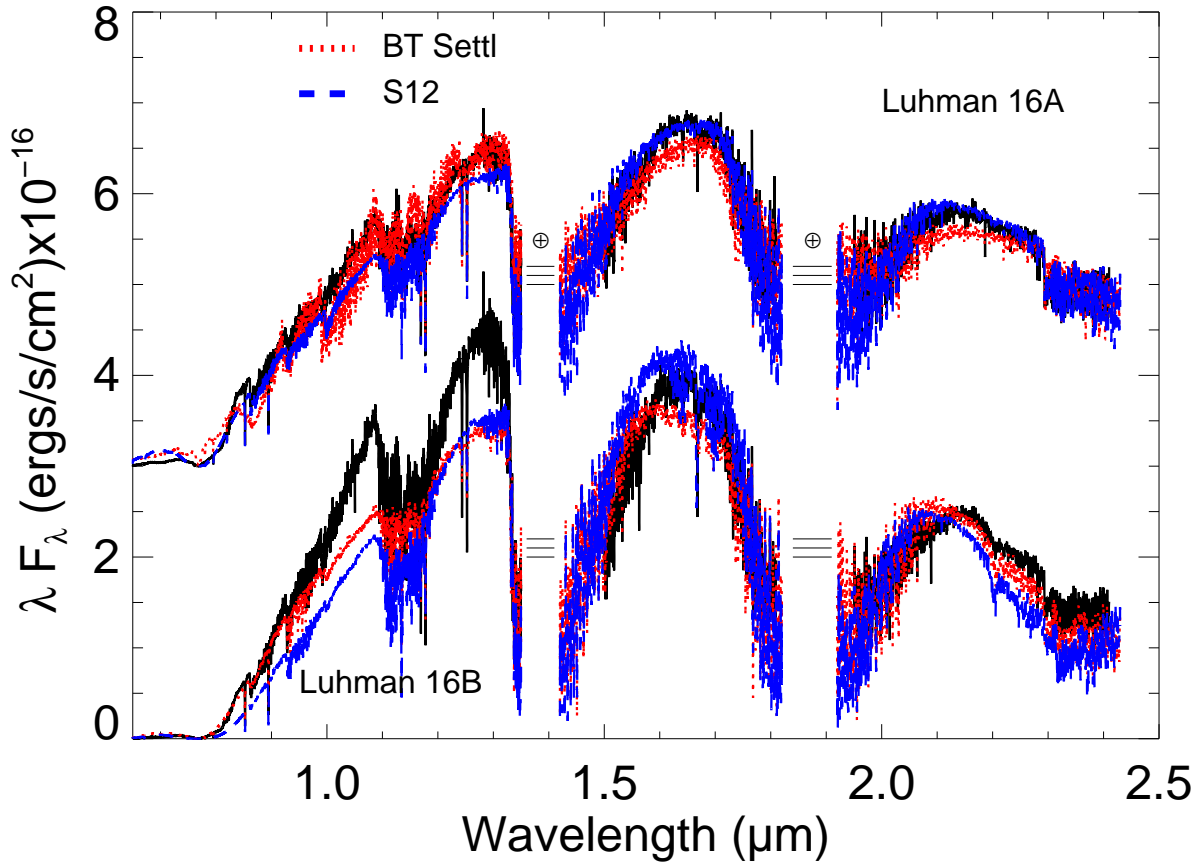}}
\caption{The near-infrared spectra scaled to the distance of the system with the model spectra corresponding to the minimum G-value (see Figure~\ref{fig:Gvalues}).  The best fit BT Settl model for Luhman 16A (top, red long-dashed) corresponds to a T$_{\mathrm eff}$=1650 and logg=5.0, and the best fit S12 (top blue short-dashed) corresponds to a T$_{\mathrm eff}$=1200, logg=5.0 that is out of chemical equilibrium.  The best fit BT Settl model for Luhman 16B (bottom, red long-dashed) corresponds to a T$_{\mathrm eff}$=1400 and logg=5.5 and the best fit S12 model (bottom blue short-dashed) corresponds to a T$_{\mathrm eff}$=900, logg=5.5 that is out of chemical equilibrium. Areas of strong telluric absorption have been removed but are marked by three horizontal lines.  The two sources are offset from one another by 3.0x10$^{-16}$ units } 
\label{fig:models}
\end{figure*}

Using the IDL $smooth$ function and $interpol$ routine we matched the spectral resolution and array size of the models to our observations, calculated the G-value for each model spectra, and examined the best fits by eye. We ignore areas of strong telluric absorption around 1.4 $\mu$m and 1.9 $\mu$m.    The model fits corresponding to the minimum G-value are over plotted in Figure~\ref{fig:models}.  

Viewing the G-value statistic over the range of model parameters in Figure~\ref{fig:Gvalues} shows that there were a number of nearly equivalent fits for both components.    To assess the uncertainty in the fitting, we performed a Monte Carlo simulation and determined the range of synthetic spectra that best fit the data given the observational errors.  The best model deduced parameter ranges are displayed in Table ~\ref{Table:Models}.

Luhman 16A is best matched to the BT Settl model with parameters of T$_{\mathrm eff}$=1650 and logg=5.0.  The $J$ band is well fit,  however the source is more luminous at both $H$ and $K$ bands.  Luhman 16B on the other hand is best matched to the BT Settl model with parameters T$_{\mathrm eff}$=1400 and logg=5.5.  Converse to the Luhman 16A fit, the secondary is more luminous at $J$ band but well fit at $H$ and $K$ bands.  

Using the S12 models, Luhman 16A is best matched with a cooler temperature of T$_{\mathrm eff}$=1400 and logg=5.0 with a cloudy photosphere that is out of chemical equilibrium.  The model $J$ band is less luminous while the $H$ and $K$ bands are well fit.  For Luhman 16B, the best fit parameters are T$_{\mathrm eff}$=900 and logg=5.5 with a cloudy photosphere that is out of chemical equilibrium.  The model $J$ is much less luminous than the data while the model $H$ band is slightly more luminous.     

Given the co-evolving nature of the system, hence the requirement that at the very least the best fit gravity and metallicity parameters should match for both components, the Luhman 16AB system will be a benchmark for calibrating atmosphere model predictions.  Unfortunately, the results here-in demonstrate that little physical information can be drawn about each component from current model comparisons alone.

\subsection{Bolometric Luminosity, T$_{eff}$, and Mass}
As discussed in Section 3, the age of the system can be constrained by the Li I absorption measurement and the lack of surface gravity features (0.1 - 3 Gyr). We can combine this age range with bolometric luminosities (L$_{bol}$) and investigate the masses of each component.  

In order to calculate L$_{bol}$, we integrated over the observed data (Mage+FIRE) supplemented with BT Settl or S12 data for longer wavelength regions (see Table ~\ref{Table:Models}).  In Table ~\ref{Table:Parameters} we report the L$_{bol}$ average from supplementing with the best fit BT Settl and best fit S12 models discussed above.  We find that the two components have consistent L$_{bol}$ values (within 1$\sigma$) therefore, as expected,  they are very close in T$_{eff}$ and mass.   

Following the prescription from \citet{Vrba04},  we calculate and report T$_{eff}$s derived from L$_{bol}$ measurements assuming a radii of 0.9 R$_{Jup}$.  These values of 1310$\pm$30 K and 1280$\pm$75 K for Luhman 16A and Luhman 16B respectively are consistent with the expected T$_{\mathrm eff}$'s for each component from the \citet{Stephens09} relations. Using the evolutionary models from \citet{Baraffe03}, we find likely masses for Luhman 16A and Luhman 16B of 20 - 40 M$_{Jup}$ at 0.5 Gyr,  30 -50 M$_{Jup}$ at 1 Gyr, and 50 - 65 M$_{Jup}$ at 3 Gyr. 

\citet{King10} find an  L$_{bol}$ value for epsilon Indi Ba of -4.699$\pm$0.017 and estimate an age of 3.7 - 4.3 Gyr based on a combination of the systems dynamical mass and evolutionary models (\citealt{Cardoso09}, \citealt{Baraffe03}).  At comparable spectral types (Luhman 16B -- T0.5, epsilon Indi Ba -- T1), temperatures, and L$_{bol}$ values we find the major difference between these two benchmark T dwarfs is the strong detection of Li I in Luhman 16B (as discussed in section 3.1).  The comparison with epsilon Indi Ba is further evidence that Luhman 16B is younger (estimated age 0.1 - 3 Gyr) and less massive ($<$ 70 M$_{Jup}$ as estimated by \citealt{King10} for epsilon IndiBa).

\section{CONCLUSIONS}
The newly discovered 2.02$\pm$0.019 pc brown dwarf binary (L7.5+T0.5) Luhman 16AB is a valuable astronomical target for low-temperature atmosphere studies.  In this work we present medium resolution optical ($\lambda$/$\Delta\lambda$$\sim$ 4000) and near-infrared ($\lambda$/$\Delta\lambda$$\sim$ 8000) data of each component in the system.  We discuss the spectral features in red optical, $zJHK$ bands highlighting prominent temperature, gravity, and atmosphere indicators among the two components.  

In the red optical we find that both components have strong 6708\,\AA\  Li I absorption confirming their status as substellar mass objects ($<$ 0.65 M$_{Jup}$) and upper age limit of $\sim$3.0~Gyr.  Interestingly this is the first Li I absorption measurement in a T dwarf.  We find strong Rb I and Cs I lines in Luhman 16A and Luhman 16B with the latter demonstrating comparably stronger equivalent widths as expected for a cooler source. In the $z$ band, we find that the FeH Wing-Ford feature, a potential signpost for atmospheric properties, is prominent and equivalent in both sources.  

The $H$ and $K$ band spectra of both components are comparable with the largest difference found at 2.2 $\mu$m where the secondary shows increased CH$_{4}$ absorption.  This is the strongest spectral indication that Luhman 16B is a later spectral type hence cooler temperature than Luhman 16A.  

In the $J$ band we find no hint of Na I absorption in either component (this is confirmed in the optical as well); however strong absorption by the K~I doublets at (1.168, 1.177) $\mu$m and (1.243, 1.254) $\mu$m.  Comparing equivalent widths of each line to a sample of late-type M, L, and T dwarfs we find that both components fall within the expected range for ultra cool dwarfs with Luhman 16A tending toward stronger lines and Luhman 16B tending toward weaker lines.  Given the close temperature range of both sources, we postulate that the stronger K~I absorption in Luhman 16B is due to thinner clouds or holes allowing us to see to deeper layers.  Examining the spectral region around each alkali doublet in detail shows that the continuum surrounding the 1.25 $\mu$m feature is brighter in Luhman 16B than Luhman 16A, confirming the flux reversal nature of this system.  This region is also regulated by condensate grain opacity therefore we interpret this as a signature of cloud variations between the two. 

Converting the flux into a brightness temperature, we find that at 1.25 $\mu$m, Luhman 16B is 50 K warmer than Luhman 16A.  At 1.17 $\mu$m, the continuum is regulated by molecular gas opacity and the brightness temperature between components is nearly equal.  We deduce that a thinner cloud layer in Luhman 16B or a patchy atmosphere revealing holes into warmer layers may explain the differences.   The corresponding near-infrared colors for each component suggest that Luhman 16A, which is significantly redder, may indeed have thicker clouds but at present shows no signs of the dynamic weather patterns seen in Luhman 16B.  

A model comparison of the near-infrared spectra of each component with the BT Settl and S12 atmospheric model synthetic spectra yields best fit temperatures of 1650 K and 1200 K for Luhman 16A respectively and 1400 K and 900 K for Luhman 16B respectively.  Investigating by eye shows that the models fit with varying levels of success.  

Using all spectral information we calculate bolometric luminosities and find near equal values for both components indicating that they must have nearly the same T$_{eff}$s (Luhman 16A 1310$\pm$30 K; Luhman 16B 1280 $\pm$ 75 K).  The resultant masses corresponding to the broad age range of 0.1 - 3 Gyr are 20 -40 M$_{Jup}$ at 0.5 Gyr, 30 - 50 M$_{Jup}$ at 1 Gyr, and 50 - 65 M$_{Jup}$ at 3 Gyr for each component.  Future dynamical mass measurements will help narrow this broad mass and age range.  

\acknowledgments{Acknowledgements }
The authors thank the anonymous referee for a very helpful and thorough report.  The authors also would like to thank M. Marley and D. Saumon for access to current atmospheric models, and E. Rice for access to NIRSPEC comparison data.   J. Faherty was supported by NSF IRFP award number 0965192 while this research was conducted. This publication uses data gathered with the 6.5 meter Magellan Telescopes located at Las Campanas Observatory, Chile and we thank the operators J. Araya, M. Gonzalez, G. Martin for assistance in acquiring data. Research has benefitted from the M, L, and T dwarf compendium housed at DwarfArchives.org and maintained by Chris Gelino, Davy Kirkpatrick, and Adam Burgasser.   This publication makes use of data products from the Two Micron All-Sky Survey, which is a joint project of the University of Massachusetts and the Infrared Processing and Analysis Center/California Institute of Technology, funded by the National Aeronautics and Space Administration and the National Science Foundation. This research has made use of the NASA/ IPAC Infrared Science Archive, which is operated by the Jet Propulsion Laboratory, California Institute of Technology, under contract with the National Aeronautics and Space Administration. 
\clearpage

\begin{table}
\caption{\scriptsize{Equivalent Widths of Prominent Optical Lines\label{Table:EQ_OPT}}}
\begin{tabular}{lccccccccccc}
\hline
\\
Component & SpT & Li  (6708\,\AA\ ) & $|$ H$\alpha$ $|$ (6563\,\AA\ )\tablenotemark{a} & Rb I (7800\,\AA\ ) & Rb I (7948\,\AA\ ) & Cs I (8521\,\AA\ ) &  Cs I (8943\,\AA\ ) \\
\hline
\\
Luhman 16A& L8.5&8.0$\pm$0.4 &$<$ 1.5 & 5.3$\pm$0.5 &6.0$\pm$0.3 &6.8$\pm$0.3 &4.0$\pm$0.3 \\
Luhman 16B& T0.5&3.8$\pm$0.4 &$<$ 1.5 & 6.2$\pm$0.5 &5.7$\pm$0.5 &7.8$\pm$0.3 &6.3$\pm$0.3 \\
\\
\hline
\tablenotetext{a}{The limit for H$\alpha$ is given as an absolute value as it applies to either emission or absorption}
\end{tabular}
\end{table}

\begin{table}
\caption{\scriptsize{Equivalent Widths of Prominent near-infrared Lines\label{Table:EQ_IR}}}
\begin{tabular}{lccccccccccc}
\hline
\\
Component & SpT & K I (1.168 $\mu$m) & K I (1.177 $\mu$m) & K I (1.243 $\mu$m) & K I (1.254 $\mu$m) \\
&&(~\AA~)&(~\AA~)&(~\AA~)&(~\AA~)\\
\hline
\\
Luhman 16A& L8.5&5.0$\pm$0.5 & 8.0$\pm$0.5 &2.7$\pm$0.2 &4.0$\pm$0.2 \\
Luhman 16B& T0.5&7.2$\pm$0.5 &11.6$\pm$0.5 &4.3$\pm$0.2 & 7.0$\pm$0.2\\
\\
\hline
\end{tabular}
\end{table}

\clearpage

\begin{table}
\caption{\scriptsize{Model Fitting Results\label{Table:Models}}}
\begin{tabular}{lccccccccccc}
\hline
\\
Component & Model & T$_{\mathrm eff}$ & logg & Metallicity& Clouds &  Chemistry \\
\hline
\\
Luhman 16A& BT Settl& 1650&5.0&0.0&---&---\\
Luhman 16A& S12& 1200&5.0&---&clouds&Out of CE\\
\hline
Luhman 16B& BT Settl& 1400&5.5&0.0&---&---\\
Luhman 16B&  S12& 900& 5.5&---&clouds&Out of CE\\

\\
\hline
\end{tabular}
\end{table}

\clearpage

\begin{table}
\caption{\scriptsize{Measured Parameters\label{Table:Parameters}}}
\begin{tabular}{lccccccccccc}
\hline
\\
 & Luhman 16A & Luhman 16B  & System & Reference\\
\hline
\\
RA (epoch 2010) & &&10 49 15.57&1\\
DEC (epoch 2010)& &&-53 19 06.1& 1\\
Distance (pc) &&&2.02$\pm$0.019 & 2\\
SpT (IR)&L7.5$\pm$0.5 & T0.5$\pm$0.5&& 3 \\
MKO J &11.53$\pm$0.04&11.22$\pm$0.04&&3\\
MKO H &10.37$\pm$0.04&10.39$\pm$0.04&&3\\
MKO K &9.44$\pm$0.07&9.73$\pm$0.09&&3\\
2MASS J\tablenotemark{a}&11.68$\pm$0.05 & 11.40$\pm$0.05&10.73$\pm$0.03&4,5\\
2MASS H\tablenotemark{a}&10.31$\pm$0.05&10.34$\pm$0.05&9.56$\pm$0.03&4,5\\
2MASS K$_{s}$\tablenotemark{a}&9.46$\pm$0.08&9.71$\pm$0.10&8.84$\pm$0.02&4,5\\
WISE W1&&&7.89$\pm$0.02&1\\
WISE W2&&&7.33$\pm$0.02&1\\
WISE W3&&&6.20$\pm$0.02&1\\
WISE W4&&&5.95$\pm$0.04&1\\
Age\tablenotemark{b}&&&0.1 - 3 Gyr& 4\\
Log(L$_{bol}$/L$_{\sun}$)&-4.67$\pm$0.04&-4.71$\pm$0.1&&4\\
T$_{eff, L_{bol}}$\tablenotemark{c}&1310$\pm$30&1280$\pm$75&&4\\

Mass\tablenotemark{d} (M$_{Jup}$ at Age 0.5 Gyr)&20 - 40 &20 - 40 &&4\\
Mass\tablenotemark{d} (M$_{Jup}$ at Age 1 Gyr)&30 - 50 &30 - 50&&4 \\
Mass\tablenotemark{d} (M$_{Jup}$ at Age 3 Gyr)&50 - 65 &50 - 65&&4 \\
\\
\hline
\end{tabular}
\tablenotetext{a} {2MASS Photometry converted from MKO values using the \cite{Stephens04} transformations}
\tablenotetext{b}{Lower age based on the lack of gravity features and upper age based on Li I absorption in both components.}
\tablenotetext{c}{T$_{eff}$ computed following the prescription in \citet{Vrba04} where the radius is assumed to be 0.9 R$_{Jup}$}
\tablenotetext{d}{Mass ranges derived using the T$_{eff}$ range of 1000 - 1400 K, the age range of 0.1 - 3 Gyr and the \cite{Baraffe03} evolutionary models}
\tablecomments{References: (1) \cite{Wright10} (2) \cite{Boffin13}  (3) \cite{Burgasser13}   (4) This work  (5) \cite{Cutri03}  }
\end{table}

\bibliographystyle{apj}
\bibliography{paper2}

\end{document}